\documentclass[fleqn,usenatbib]{mnras}

\usepackage{newtxtext,newtxmath}

\usepackage[T1]{fontenc}

\DeclareRobustCommand{\VAN}[3]{#2}
\let\VANthebibliography\thebibliography
\def\thebibliography{\DeclareRobustCommand{\VAN}[3]{##3}\VANthebibliography}

\usepackage{graphicx}	
\usepackage{amsmath}	
\usepackage{makecell}
\usepackage{bm}

\title[Full-shape analysis of BOSS \& eBOSS]{Cosmological implications of the full shape of 
anisotropic clustering measurements in BOSS and eBOSS}

\author[A. Semenaite et al.]{
Agne Semenaite$^{1}$\thanks{E-mail: agne@mpe.mpg.de},
Ariel G. S\'anchez$^{1,2}$,
Andrea Pezzotta$^{1}$,
Jiamin Hou$^{1,3}$,
Roman Scoccimarro$^{4}$, 
\newauthor 
Alexander Eggemeier$^{5}$,
Martin Crocce$^{6,7}$, 
Chia-Hsun Chuang$^{8}$,
Alexander Smith$^{9}$,
Cheng Zhao$^{10}$,
\newauthor
Joel R. Brownstein$^{11}$,
Graziano Rossi$^{12}$
and Donald P. Schneider$^{13, 14}$
\\
$^{1}$ Max-Planck-Institut f\"ur extraterrestrische Physik, Postfach 1312, Giessenbachstr., 85748 Garching, Germany\\ 
$^{2}$Universit\"as-Sternwarte M\"uchen, Scheinerstrasse 1, D-81679 M\"uchen, Germany\\
$^{3}$Department of Astronomy, University of Florida,
211 Bryant Space Science Center, Gainesville, FL 32611, USA\\
$^{4}$ Center for Cosmology and Particle Physics, Department of Physics, New York University, NY 10003, New York, USA\\
$^{5}$  Institute for Computational Cosmology, Department of Physics, Durham University, South Road, Durham DH1 3LE, United Kingdom\\
$^{6}$  Institute of Space Sciences (ICE, CSIC), Campus UAB, Carrer de Can Magrans, s/n, 08193 Barcelona, Spain\\
$^{7}$  Institut d’Estudis Espacials de Catalunya (IEEC), 08034 Barcelona, Spain\\
$^{8}$ Kavli Institute for Particle Astrophysics and Cosmology, Stanford University, 452 Lomita Mall, Stanford, CA 94305, USA\\
$^{9}$ IRFU, CEA, Université Paris-Saclay, F-91191 Gif-sur-Yvette, France\\
$^{10}$Institute of Physics, Laboratory of Astrophysics, \'Ecole Polytechnique F\'ed\'erale de Lausanne (EPFL), Observatoire de Sauverny, CH-1290 Versoix, Switzerland\\
$^{11}$ Department of Physics and Astronomy, University of Utah, 115 S. 1400 E., Salt Lake City, UT 84112, USA\\
$^{12}$ Department of Physics and Astronomy, Sejong University, Seoul, 143-747, Republic of Korea\\
$^{13}$ Department of Astronomy and Astrophysics, The Pennsylvania State University, University Park, PA 16802\\
$^{14}$ Institute for Gravitation and the Cosmos, The Pennsylvania State University, University Park, PA 16802
}
\date{Accepted XXX. Received YYY; in original form ZZZ}

\pubyear{2021}

\begin{document}
\label{firstpage}
\pagerange{\pageref{firstpage}--\pageref{lastpage}}
\maketitle

\begin{abstract}
We present the analysis of the full shape of anisotropic clustering measurement from the 
extended Baryon Oscillation Spectroscopic Survey (eBOSS) quasar sample together with 
the combined galaxy sample from the Baryon Oscillation Spectroscopic Survey (BOSS), 
re-analysed using an updated recipe for the non-linear matter power spectrum and the 
non-local bias parameters. We obtain constraints for flat $\Lambda$CDM cosmologies, focusing on 
the cosmological parameters that are independent of the Hubble parameter $h$. Our recovered 
value for the RMS linear perturbation theory variance as measured on the scale of $12\,{\rm Mpc}$ 
is $\sigma_{12}=0.805\pm 0.049$, while using the traditional reference scale of 
$8\,h^{-1}{\rm Mpc}$ gives $\sigma_{8}=0.815\pm 0.044$. We quantify the agreement between 
our measurements and the latest CMB data from {\it Planck} 
using the suspiciousness metric, and find them to be consistent within $0.64 \pm 0.03\sigma$. 
Combining our clustering constraints with the $3\times2$pt data sample from 
the Dark Energy Survey (DES) Year 1 release slightly degrades this agreement to the level of $1.54 \pm 0.08\sigma$, while still showing an overall consistency with \textit{Planck}. 
We furthermore study the effect of imposing a {\it Planck} - like prior on the parameters that define the 
shape of the linear matter power spectrum, and find significantly tighter 
constraints on the parameters that control the evolution of density fluctuations. In particular, 
the combination of low-redshift data sets prefers a value of the physical dark energy density 
$\omega_{\rm DE}=0.335 \pm 0.011$, which is 1.7$\sigma$ higher than the one preferred by 
{\it Planck}.
\end{abstract}




\section{Introduction}

The rapid progress of observational cosmology in recent years has been fuelled by an abundance 
of accurate observations \citep{Riess1998, Perlmutter1999, Eisenstein2005, Cole2005, wmap9, 
Anderson2012, Alam17,Planck2018,eboss2021}. The $\Lambda$CDM model has emerged 
as the new cosmological paradigm, being able to simultaneously describe all state-of-the-art observations. 
However, the two components making up the majority of the total energy budget of the Universe 
today in this model, dark energy and dark matter, remain poorly understood.

As it provides the most precise parameter constraints to date, the best-fitting 
$\Lambda$CDM model to the cosmic microwave background (CMB) observations by the 
\textit{Planck} satellite \citep{Planck2018} has become the synonym to `standard cosmological model'.
The comparison of predictions for the expansion history of the Universe $H(z)$, and the rate at which 
cosmic structures form at later times $f(z)$ with observations at lower redshifts 
serves as a powerful test of the $\Lambda$CDM paradigm. 
While a broad range of observations are in agreement with the CMB predictions, 
the increasingly precise measurements from the cosmic distance ladder \citep{Riess2018,Riess_2019}, 
as well as the ever-increasing weak gravitational lensing data sets 
\citep{Hildebrandt_2016, Abbott2017b, Hikage2019}, display hints of tension with Planck $\Lambda$CDM. 
In particular, local direct probes seem to prefer an up to 5$\sigma$ \citep{Riess2021} greater expansion rate of the Universe (the `$H_0$ tension') while, less significantly \citep[up to 3$\sigma$,][]{kids1000}, weak gravitational lensing measurements exhibit
a lower amplitude  (the `$\sigma_8$ tension') than predicted by Planck. There 
is still no conclusive evidence on what drives these differences in the recovered values, and 
it is not uncommon for proposed solutions to the $H_0$ tension to make the $\sigma_8$ tension 
worse \citep[see, for example,][]{Hill_2020}.

With the lack of consensus on where (and whether) the inconsistencies with Planck $\Lambda$CDM 
arise, the cosmological community has put in an increased effort in assessing the internal consistency 
between different data sets within the $\Lambda$CDM scenario. Galaxy clustering allows to probe and 
distinguish how the underlying cosmology affects the background expansion of the Universe and its 
effects on the structure growth through baryon acoustic oscillations (BAO) and redshift-space 
distortions (RSD). Both of these effects set important features of the two-point correlation function, 
which can be fit to obtain summary statistics that carry compressed cosmological information. 
BAO set the angular scale of the acoustic peak, allowing to probe the distance -- redshift relation. 
On the other hand, RSD provide information about structure growth through galaxy peculiar 
velocities, whose effect on the amplitude of the power spectrum is commonly characterised by 
the product of the logarithmic growth rate $f$ and the RMS linear perturbation theory variance 
$\sigma_8$ \citep[although see][for the problems caused by this approach]{sanchez2020let}. 
While analyses based solely on RSD and BAO summary statistics allow excellent internal 
consistency tests and may help constrain beyond - $\Lambda$CDM scenarios, it has been shown that 
they do not preserve all the information of the full measurement, in particular, losing the additional constraining power available from its shape \citep{ShapeFit}.

Other analyses, therefore, make use of the information recovered from fitting the full shape of two-point
clustering measurements, either in Fourier or configuration space, directly comparing models against 
data \citep{Tr_ster_2020, dAmico_2020, Ivanov_2020, Chen2021}. These analyses tend to lose some of 
the immediate interpretability of the summary statistics but instead allow to directly obtain 
constraints of cosmological parameters independently of external data sets. This type of 
analyses have, therefore, recently received attention as a way to test the consistency 
between large-scale structure (LSS) and CMB measurements. 

\citet{Tr_ster_2020} showed that the full shape analysis of galaxy clustering produces cosmological constraints that are comparable to those of other low-redshift probes. This work followed the analysis of correlation function wedges of Baryon Oscillation Spectroscopic Survey (BOSS) galaxies by \citet{S17}  in order to derive constraints on flat $\Lambda$CDM cosmologies from galaxy clustering alone \citep[i.e. without combining it with CMB measurements, as was done in the BOSS Data Release 12 consensus analysis,][] {Alam17}. Furthermore, the work also presented joint low-redshift constraints by combining galaxy clustering with weak lensing measurements from the Kilo-Degree Survey (KV450).
 
The $\sigma_8$ value recovered from the full shape analysis of the correlation function wedges by \citet{Tr_ster_2020} is 2.1$\sigma$ low  compared to Planck's 
prediction, with the difference increasing to 3.4$\sigma$ when weak lensing measurements from KV450, are added, indicating that there may be some consistent discrepancy 
between CMB predictions and low redshift observations. This is also consistent with the more 
recent analysis by \citet{kids1000} where BOSS galaxies are used as lenses in the so 
called `$3\times2$pt' analysis (a set of three correlation functions consisting of autocorrelation of the lens 
galaxy positions, source galaxy shapes and the cross-correlation of the two), which finds a 
$\sim3\sigma$ discrepancy with \textit{Planck}'s value of $S_8=\sigma_8\sqrt{\Omega_m/0.3}$, that 
combines $\sigma_8$ with the matter density $\Omega_m$ in a way that minimises correlation 
between the two parameters. This result is consistent with the findings from other major weak 
lensing surveys (Dark Energy Survey \citep{desyr3} and Hyper Suprime-Cam \citep{Hikage2019}), 
even though, most recently, DES reported consistency between their $3\times2$pt $\Lambda$CDM 
constraints and those of Planck when the full parameter space is considered. Furthermore, 
$\sigma_8$ may not be an entirely appropriate parameter for assessing consistency among the 
different surveys, as \citet{sanchez2020let} have shown that it is affected by the different posterior 
distributions of Hubble parameter $h$ recovered by different analyses. Alternatively, one may define 
$\sigma_{12}$ - the variance as measured on a fixed scale of 12 Mpc. In this work we adopt this 
notation and use $\sigma_{12}$ to both accurately characterise the amplitude of the power spectrum 
today as well as assess the consistency among the probes considered. 

In this work we are, therefore, interested in building upon \citet{Tr_ster_2020} and exploring, 
whether the discrepancy between the low-redshift probes and Planck within the $\Lambda$CDM 
model holds when extending the redshift range probed by the clustering measurements with the 
addition of eBOSS quasar clustering. We provide the joint constraints from the full shape analysis 
of BOSS galaxy and eBOSS quasar clustering on their own, as well as in combination with weak 
lensing information. For our weak lensing data set we use the $3\times 2$pt measurements from the 
Dark Energy Survey Year 1 \citep[DES Y1,][]{Abbott2017b} release, which both cover a larger area 
than KV450 and include 
galaxy clustering and galaxy-galaxy lensing as well as the shear-only measurements. If the tension 
seen between the low-redshift probes and Planck is purely statistical, adding more data should not 
only tighten the posterior contours but be able to bring the constraints to a better agreement. The 
results from an equivalent analysis with KiDS-450 shear measurements are available in the 
appendix \ref{appendix:kids}. 

In addition to expanding our data sets we also aim to re-define the parameter space following 
\citet{Sanchez2021}, who distinguish `shape' and `evolution' cosmological parameters. This classification is introduced to describe the degenerate way in which evolution parameters affect the linear matter power spectrum when expressed in Mpc units.
In such parameter space, $\sigma_8$ is replaced by $\sigma_{12}$, as discussed above, 
and the relative matter and dark energy densities ($\Omega_{\rm{m}}$, $\Omega_{\rm{DE}}$) are 
replaced by their physical counterparts ($\omega_{\rm{m}}$, $\omega_{\rm{DE}}$). While in \citet{Sanchez2021} the $h$-independent parameter space is presented to create a framework which allows them to reduce the number of parameters required to model the cosmology dependence of the matter power spectrum, the advantage of such parameter choice for this work is two-fold. First, the derived constraints do not depend on the 
posterior of $h$ of the particular analysis and can, therefore, be directly compared with 
constraints from other data sets and, second, the effect that each of the parameters has 
on the power spectrum is clear, with evolution parameters affecting its amplitude and 
shape parameters determining the shape. 

We provide a more detailed description of the parameter space we use (including the 
prior choices) in Section \ref{section:methods}, together with a summary of our data and models, 
and illustrate how it compares with its $h$-dependent equivalent in Section \ref{sec:clustering}, 
where we also present our cosmological constraints \textbf{from} BOSS and eBOSS. The results 
obtained when adding DES are further presented in Section \ref{sec:joint} We finish with a 
discussion of our results in Section \ref{section:discussion} and present our conclusions 
in Section \ref{section:conclusions}. 

\section{Methodology}
\label{section:methods}

This work is an extension of \citet{Tr_ster_2020} and largely follows the 
same structure and methods - we assume 
flat $\Lambda$CDM cosmology and obtain the joint low-redshift parameter 
constraints by combining the likelihoods 
for each data set considered independently. Our model for anisotropic galaxy 
and quasar clustering measurements 
follows that described in \citet{S17} for the so-called `full shape' analysis 
(with the exception of the matter power spectrum model) whereas for the 
`$3\times 2$pt' analysis (galaxy shear, galaxy-galaxy lensing, and galaxy clustering)  
we use the model described in \citet{Abbott2017b}. In this section we summarise 
the data and models used with a more detailed description available in the references above. 
The measurements here are as used in the respective original analyses and, therefore, had 
been tested against various systematics and include the appropriate corrections. 

\subsection{Galaxy and QSO clustering measurements}
\label{sec:data}

The Sloan Digital Sky Survey (SDSS) has mapped the large-scale structure of the Universe 
thanks to the accurate measurements by the double-armed spectrographs \citep{SDSS_spectro}
at the Sloan Foundation Telescope at Apache Point Observatory \citep{SDSS_telescope}. 
Throughout its different stages \citep{York2000, SDSS_III, SDSS_IV} the SDSS has provided 
redshift information on millions of galaxies and quasars. 

We consider clustering measurements in configuration space from two data sets: 
the galaxy samples of BOSS \citep{BOSS}, corresponding to SDSS DR12 \citep{Alam2015, Reid2016}, 
and the QSO catalogue \citep{Lyke2020} from eBOSS \citep{eBOSS}, contained in SDSS DR16 \citep{sdss_dr16,ross2020}. 
In each case, the information from the full anisotropic correlation function
$\xi(s,\mu)$,  where $s$ denotes the comoving pair separation and
$\mu$ represents the cosine of the angle between the separation vector and the line of sight, was compressed into different but closely related statistics.

We analyse the clustering properties of the combined BOSS galaxy sample using the 
measurements of \citet{S17}, who employs the clustering wedges statistic \citep{Kazin2012}, 
$\xi_{\Delta\mu}(s)$, which corresponds to the average of $\xi(s,\mu)$, over the interval 
$\Delta\mu=\mu_{2}-\mu_{1}$, that is
\begin{equation}
\xi_{\Delta\mu}(s)= \frac{1}{\Delta \mu}\int^{\mu_2}_{\mu_1}{\xi(\mu,s)}\,{{\rm d}\mu}.
\label{eq:wedges}
\end{equation}
\citet{S17} measured three wedges by splitting the $\mu$ range from 0 to 1 into 
three equal-width intervals. We consider their measurements in two redshift bins, with 
$0.2 < z < 0.5$ (the LOWZ sample) and $0.5 < z < 0.75$ (CMASS), corresponding to the effective redshifts
$z_{\rm eff} = 0.38$ and 0.61, respectively. 
The covariance matrices, $\mathbfss{C}$, of these data were estimated using the set of 2045 
{\sc MD-Patchy} mock catalogues described in \citet{Kitaura2016}.
These measurements were also used in the analysis of \citet{Tr_ster_2020} and the 
recent studies of the cosmological implications of the KiDS 1000 data set \citep{kids1000, troster2021}.

For the eBOSS QSO catalogue we use the measurements of \citet{Hou2021}, who considered 
the Legendre multipoles given by
\begin{equation}
\xi_{\ell}(s) =\frac{2\ell+1}{2} \int^{1}_{-1} \xi(\mu,s) \mathcal{L}_{\ell}(\mu) \,{\rm d}\mu,
\label{eqn:multipoles}
\end{equation}
where $\mathcal{L}_{\ell}(\mu)$ denotes the $\ell$-th order Legendre polynomial.
We consider the multipoles $\ell = 0,\,  2, \, 4$ obtained using the redshift range
$0.8 < z < 2.2$, with an effective redshift $z_{\rm eff} = 1.48$.
The covariance matrix of these measurements were obtained using 
the set of 1000 mock catalogues described in \citet{Zhao2021}. 
Besides the QSO sample used here, the full 
eBOSS data set contains two additional tracers, the luminous red galaxies (LRG)
and emission line galaxies (ELG) samples
\citep[for the corresponding  BAO and RSD analyses, see][]{Bautista2021, GilMarin2020, 
deMattia2021,Tamone2020}.
These samples overlap in redshift among them and with the galaxies from BOSS.
We therefore restrict our analysis of eBOSS data to the QSO sample to, in combination with BOSS, 
cover the maximum possible redshift range while ensuring that the clustering measurements can 
be treated as independent in our likelihood analysis.

We treat the measurements from BOSS and eBOSS as in the original analyses of \citet{S17} and 
\citet{Hou2021}. We restrict our analysis to pair separations within the range
$20\,h^{-1}{\rm Mpc} < s< 160\,h^{-1}{\rm Mpc}$. We assume a Gaussian likelihood 
for each set of measurements, in which the covariance matrices are kept fixed. 
We account for the impact of the finite number of mock catalogues used 
to derive $\mathbfss{C}$ \citep{kaufman1967,Hartlap2007,Percival2014}.
The large number of mock catalogues used ensures that the effect of the 
noise in $\mathbfss{C}$ on the obtained cosmological constraints corresponds to a modest 
correction factor of less than 2 per cent.

\subsection{Modelling anisotropic clustering measurements}
\label{section:model}

Our modelling of the full shape of the  Legendre multipoles and clustering wedges of the anisotropic 
two-point correlation function largely follows the treatment of \citet{S17}, with some 
important differences. 

We compute model predictions of the non-linear matter power spectrum, $P_{\rm{mm}}(k)$, 
using the Rapid and Efficient SPectrum 
calculation based on RESponSe functiOn approach \citep[{\sc respresso},][]{respresso}. 
The key ingredient of {\sc respresso} is the response function, $K(k,q)$, which 
quantifies the variation of the non-linear matter power spectrum at scale $k$ induced by a change 
of the linear power at scale $q$ as
\begin{equation}
K(k,q)\equiv q\frac{\partial P_{\rm{mm}}(k)}{\partial P_{\rm{L}}(q)}.
\end{equation}
\cite{NisBerTar1712} presented a phenomenological model for $K(k,q)$ based on renormalised 
perturbation theory \citep{regpt}, which gives a good agreement with simulation results 
over a wide range of scales for $k$ and $q$. 
The response function allows to obtain $P_{\rm{mm}}(k)$ for arbitrary cosmological parameters 
$\pmb{\theta}$ based on a measurement from N-body simulations of a fiducial cosmology 
$\pmb{\theta}_{\rm{fid}}$ as
\begin{equation}
\begin{split}
P_{\rm{mm}}(k|\pmb{\theta}) &= P_{\rm{mm}}(k|\pmb{\theta}_{\rm{fid}})\int {\rm d}\ln q\,K(k,q)\\
&\times [P_{\rm{L}}(q|\pmb{\theta})-P_{\rm{L}}(q|\pmb{\theta}_{\rm{fid}})].
\end{split}
\label{eq:resp}
\end{equation}
The choice of $\pmb{\theta}_{\rm{fid}}$ in {\sc respresso} corresponds to the best-fitting 
$\Lambda$CDM model to the 
\textit{Planck} 2015 data \citep{Planck2015}. Equation (\ref{eq:resp}) is most accurate for cosmologies 
that are close to $\pmb{\theta}_{\rm{fid}}$. For cosmologies further away from the 
fiducial, its accuracy can be improved by performing a multi-step reconstruction. 
\citet{biasmodel}  showed that {\sc respresso} 
outperforms other perturbation theory based models 
in terms of the range of validity and accurate recovery of mean posterior values. 

Following the notation of \citet{Eggemeier2019}, we describe the relation between the galaxy density 
fluctuations, $\delta$, and the matter density fluctuations, $\delta_{\rm m}$, at one loop in terms of 
the four-parameter model
\begin{equation}
    \delta = b_1\delta_m+\frac{b_2}{2}\delta_m^2+\gamma_2\mathcal{G}_2(\Phi_v)+\gamma_{21}\mathcal{G}_2(\varphi_1, \varphi_2)+... ,
\end{equation}
where the first two terms represent contributions from linear and quadratic local bias,  while the 
remaining ones correspond to non-local terms.
Here, $\mathcal{G}_2$ is the Galileon operator of the normalized velocity potential $\Phi_{\nu}$, and 
$\varphi_1$ is the linear Lagrangian perturbation potential with $\varphi_2$ as a second-order 
potential that accounts for the non-locality of the gravitational evolution,
\begin{align}
    \mathcal{G}_2(\Phi_{\nu})&=(\nabla_{ij}\Phi_{\nu})^2 - (\nabla^2\Phi_{\nu})^2,\\
    \mathcal{G}_2(\varphi_1, \varphi_2)&=\nabla_{ij}\varphi_2\nabla_{ij}\varphi_1 - \nabla^2\varphi_{2}\nabla^2\varphi_{1}.
\end{align}
Two-point statistics alone do not constrain $\gamma_2$  well, because $\gamma_2$ enters at higher order and is degenerate with $\gamma_{21}$. Therefore, we set the value 
of this parameter in terms of the linear bias $b_1$ using the quadratic relation
\begin{equation}
\gamma_{2}(b_1) = 0.524-0.547b_1+0.046b_1^2,
\label{eq:gamma2}
\end{equation}
which describes the results of \citet*{tidal1} using excursion set theory. 
\citet{biasmodel} showed that this relation is more accurate for tracers with $b_1\gtrsim 1.3$
than the one obtained under the assumption of local bias in Lagrangian space used in \citet{S17}. 
 
The value of $\gamma_{21}$ can also be derived in terms of $b_1$ under the 
assumption of the conserved evolution of 
galaxies (hereafter co-evolution) after their formation 
as \citep{Fry9604,CatLucMat9807,CatPorKam0011,Chan2012}
\begin{equation}
    \gamma_{21}= - \frac{2}{21}(b_1-1)+\frac{6}{7}\gamma_2.
\label{eq:coevol} 
\end{equation}
This relation was thoroughly tested against constraints derived from a combination of power spectrum and bispectrum data in \citet{Eggemeier2021}, and found to be in excellent agreement for BOSS galaxies. In addition to this, in Sec.~\ref{section:validation}, we confirm that the use of this relation gives an accurate description 
of the results of N-body simulations and we therefore implement it in our analysis of the BOSS and 
eBOSS data. In this way, the only required free bias parameters in our recipe are $b_1$ and $b_2$, while the non-local bias terms can be fully expressed in terms of the  linear bias through 
equations (\ref{eq:gamma2}) and (\ref{eq:coevol}). 

Our description of the effects of RSD matches that of \citet{S17}. 
Following \citet{RSD} and \citet{Taruya_2010}, we write the two dimensional 
redshift-space power spectrum as
\begin{equation}
P(k,\mu) = W_\infty(i f k \mu) \, P_{\rm novir}(k,\mu),
\label{Prsd}
\end{equation}
where the `no-virial' power spectrum, $P_{\rm novir}(k,\mu)$, is computed using the 
one-loop approximation and includes three terms, one representing a non-linear version of the 
Kaiser formula \citep{Kaiser} 
and two higher-order terms that include the contributions of the cross-spectrum and 
bispectrum between densities and velocities.
Besides the non-linear matter power spectrum, $P_{\rm novir}(k,\mu)$ requires also the 
the velocity-velocity and matter-velocity power spectra, which we compute using the empirical 
relations measured from N-body simulations of \citep{Bel2019}. 
The function $W_{\infty}(\lambda=ifk\mu)$ represents the 
large-scale limit of the generating function of the pairwise velocity distribution, which accounts for 
non-linear corrections due to fingers-of-God (FOG) or virial motions and can be parametrised 
as \citep{S17}
\begin{equation}
W_{\infty}(\lambda)=\frac{1}{\sqrt{1-\lambda^2a^2_{\rm{vir}}}}\,\exp\left(\frac{\lambda^2\sigma^2_v}{1-\lambda^2\mathnormal{a}^2_{\rm{vir}}}\right),
\end{equation}
where $a_{\rm{vir}}$ is a free parameter characterizing the kurtosis of the small-scale velocity 
distribution, and $\sigma_{\rm{v}}$ is the one-dimensional linear velocity dispersion defined 
in terms of the linear matter power spectrum as 
\begin{equation}
 \sigma_v^2 \equiv \frac{1}{6\pi^2}\int {\rm d}k\,P_{\rm L}(k).
\label{eq:sigmav}
\end{equation}

The QSO sample is known to be affected by non-negligible redshift errors which also affect the clustering measurements \citep{Zarrouk2018}. We account for this
following \citet{Hou_2018}, who showed that this effect can be correctly described by 
including an additional damping factor to the power spectrum of equation~(\ref{Prsd}) 
of the form $\exp\left(-k\mu \sigma_{\rm err}\right)$, where $\sigma_{\rm{err}}$
is treated as an additional free parameter. 

Finally, the Alcock-Paczynski distortions \citep{Alcock1979} due to the difference 
between the true and fiducial cosmologies 
are accounted for by introducing the geometric distortion factors 
\begin{align}
q_{\bot} &=D_{\rm{M}}(z_{\rm eff})/D_{\rm{M}}'(z_{\rm eff}),\\
q_{\parallel} &=H'(z_{\rm eff})/H(z_{\rm eff}). 
\end{align}
Here, $D_{\rm{M}}(z)$ is the comoving angular diameter distance and $H(z)$ is the Hubble parameter,  
with primed quantities corresponding to the fiducial cosmology used to convert redshifts to distances. 
The distortion factors are then applied to rescale the separations $s$ of galaxy pairs and the angles 
between the separation vector and the line of sight $\mu$ such that 
\begin{align}
s &=s'\left( q_{\parallel}^2\mu'^2+q^2_{\bot}(1-\mu'^2)\right),\\
\mu &=\mu'\frac{q_\parallel}{\sqrt{q_{\parallel}^2\mu'^2+q^2_{\bot}(1-\mu'^2)}}.
\end{align}

In summary, our model of the clustering wedges from BOSS requires three free parameters, 
$b_1$, $b_2$, and $a_{\rm vir}$, with the values of $\gamma_2$ and $\gamma_{21}$ given 
in terms of $b_1$ using equations~(\ref{eq:gamma2}) and (\ref{eq:coevol}). This is one less 
free parameter than in the original analysis of \citet{S17}. The Legendre multipoles of the eBOSS 
QSO require the addition of $\sigma_{\rm{err}}$, leading to a total of four free parameters .

\subsection{Additional data sets}

We complement the information from our clustering measurements with the 
$3\times2$pt measurements from DES Y1 \citep{Abbott2017b}. 
We also use the shear measurements from the Kilo-Degree Survey \citep[KiDS-450,][]{Hildebrandt_2016} 
and present the results in  Appendix~\ref{appendix:kids}.

The source galaxy samples from DES are split into four redshift bins, spanning the redshift range of 
$0.2<z\leq1.3$. In addition to shear measurements from the source galaxies, the DES Y1 data set 
also includes galaxy clustering and galaxy-galaxy lensing two-point correlation function measurements, 
as well as the lens redshift distributions for five redshift bins in the range of $0.15<z<0.9$. 
Our scale cuts for these measurements match those of  \citet{Abbott2017b}. 

For our $3\times 2$pt analysis, we use the DES likelihood as implemented in \textsc{CosmoMC} 
\citep{Lewis_2002}, which corresponds to the model described in \citet{Abbott2017b}. The likelihood includes 
models for the two-point correlation functions describing galaxy-galaxy lensing, galaxy 
clustering, and cosmic shear. The correlation functions are modelled making use of Limber and 
flat-sky approximations \citep{Limber1, Limber2, Limber3, Limber4} with the non-linear power 
spectrum obtained using \textsc{HMCode} \citep{HMcode}
as implemented in \textsc{camb} \citep{Lewis_2000}. 
The smallest angular separations considered correspond to a comoving scale of $8\,h^{-1}{\rm Mpc}$. 
The intrinsic alignment is modelled using a `non-linear linear' 
alignment recipe \citep{IA1, IA2}. 
The model also includes a treatment for multiplicative shear bias and photometric redshift uncertainty. 
The former is accounted for by introducing multiplicative bias terms of the form $(1+m^i)$ for 
each bin $i$ for shear and galaxy-galaxy lensing. The latter is modelled by the shift parameters 
$\delta z^i$ assigned to each bin for both source and lens galaxies. Finally, baryonic effects are 
not included as they are expected to be below the measurement errors for the range of scales 
considered in the analysis. For all the weak lensing nuisance parameters we impose the same 
priors as the ones listed in \citet{Abbott2017b}.

Additionally, we test the consistency of the low-redshift LSS measurements with 
the latest CMB temperature and polarization power spectra from the \textit{Planck} 
satellite \citep{Planck2018}, to which we refer simply as `Planck'. We do not include CMB lensing information. 
We use the public nuisance parameter-marginalised  
likelihood \texttt{plik\_lite\_TTTEEE+lowl+lowE} for all Planck constraints \citep{Planck2018}.

\begin{table}
	\centering
	\caption{Priors used in our analysis. $U$ indicates a flat uniform prior within the specified range. 
The priors on the cosmological and clustering nuisance parameters match those of 
\citet{Tr_ster_2020} with the exception of $n_{\rm s}$, for which the allowed range is widened. 
The priors on the nuisance parameters of weak lensing data sets match those 
of \citet{Abbott2017b}.}
	\label{tab:priors}
	\begin{tabular}{cc} 
		\hline
		Parameter & Prior \\
		\hline
		\hline
		\multicolumn{2}{c}{Cosmological parameters} \\
		\hline
		$\Omega_{\rm{b}}h^2$ & $U(0.019, 0.026)$ \\
		$\Omega_{\rm{c}}h^2$ & $U(0.01, 0.2)$ \\
		100$\theta_{\rm{MC}}$ & $U(0.5, 10.0)$ \\
		$\tau$ & $U(0.01, 0.8)$ \\
		$\rm{ln}( 10^{10}A_{\rm{s}})$ & $U(1.5, 4.0)$ \\
		$\rm{n}_s$ & $U(0.5, 1.5)$ \\
		\hline
		\multicolumn{2}{c}{Clustering nuisance parameters} \\
		\hline
        $b_1$ & $U(0.5, 9.0)$ \\
        $b_2$ & $U(-4.0, 8.0)$ \\
        $a_{\text{vir}}$ & $U(0.0, 12.0)$ \\
        $\sigma_{\rm err}\text{ (eBOSS only)}$ & $U(0.01, 6.0)$ \\
        \hline
		
	\end{tabular}
\end{table}

\subsection{Parameter spaces and prior ranges}
\label{sec:parameters}

Our goal is to obtain constraints on the parameters of the standard $\Lambda$CDM 
model, which corresponds to a flat universe, where dark energy is 
characterized by a constant equation of state parameter $w_{\rm DE} = -1$. 
Following \citet{Sanchez2021}, we focus on cosmological parameters that can be classified as either ``shape'' or ``evolution''.
The former are parameters that control the shape of the linear-theory power spectrum expressed 
in Mpc units. The latter only affect the amplitude of $P_{\rm L}(k)$ at any given redshift.
Assuming a fixed total neutrino mass of $\sum m_{\nu} =0.06\,{\rm eV}$, the $\Lambda$CDM model can be described by the parameters
\begin{equation}
\pmb{\theta} = \left(\omega_{\rm b},\omega_{\rm c},\omega_{\rm DE}, A_{\rm s}, n_{\rm s} \right).
\label{eq:base_lcdm}
\end{equation}
These are the present-day physical energy densities of baryons, cold dark matter, and
dark energy, and the amplitude and spectral index of the primordial power spectrum 
of scalar perturbations at the pivot wavenumber of $k_0= 0.05\,{\rm Mpc}^{-1}$.

Additional parameters can be derived from the set of equation~(\ref{eq:base_lcdm}). 
The dimensionless Hubble parameter, $h$,  is defined by the sum of all 
energy contributions. For a $\Lambda$CDM model, this is
\begin{equation}
h^2 = \omega_{\rm b} + \omega_{\rm c} + \omega_{\nu} + \omega_{\rm DE}.
\label{eq:hubble}
\end{equation}
It is also common to express the contributions of the various energy components in terms 
of the density parameters 
\begin{equation}
\Omega_i = \omega_i/h^2,
\label{eq:Omegas}
\end{equation}
which represent the fraction of the total energy density of the Universe corresponding to
a given component $i$. The overall amplitude of matter density fluctuations is often 
characterized in terms of $\sigma_{8}$, the linear-theory RMS mass fluctuations in spheres 
of radius $R=8\,h^{-1}{\rm Mpc}$. 
A common property of these parameters is their dependence on the value of $h$. The issues associated with this dependence are discussed in detail by \citet{sanchez2020let} 
and can be summarised as follows.

\begin{figure}
\centering
\includegraphics[width=0.95\columnwidth]{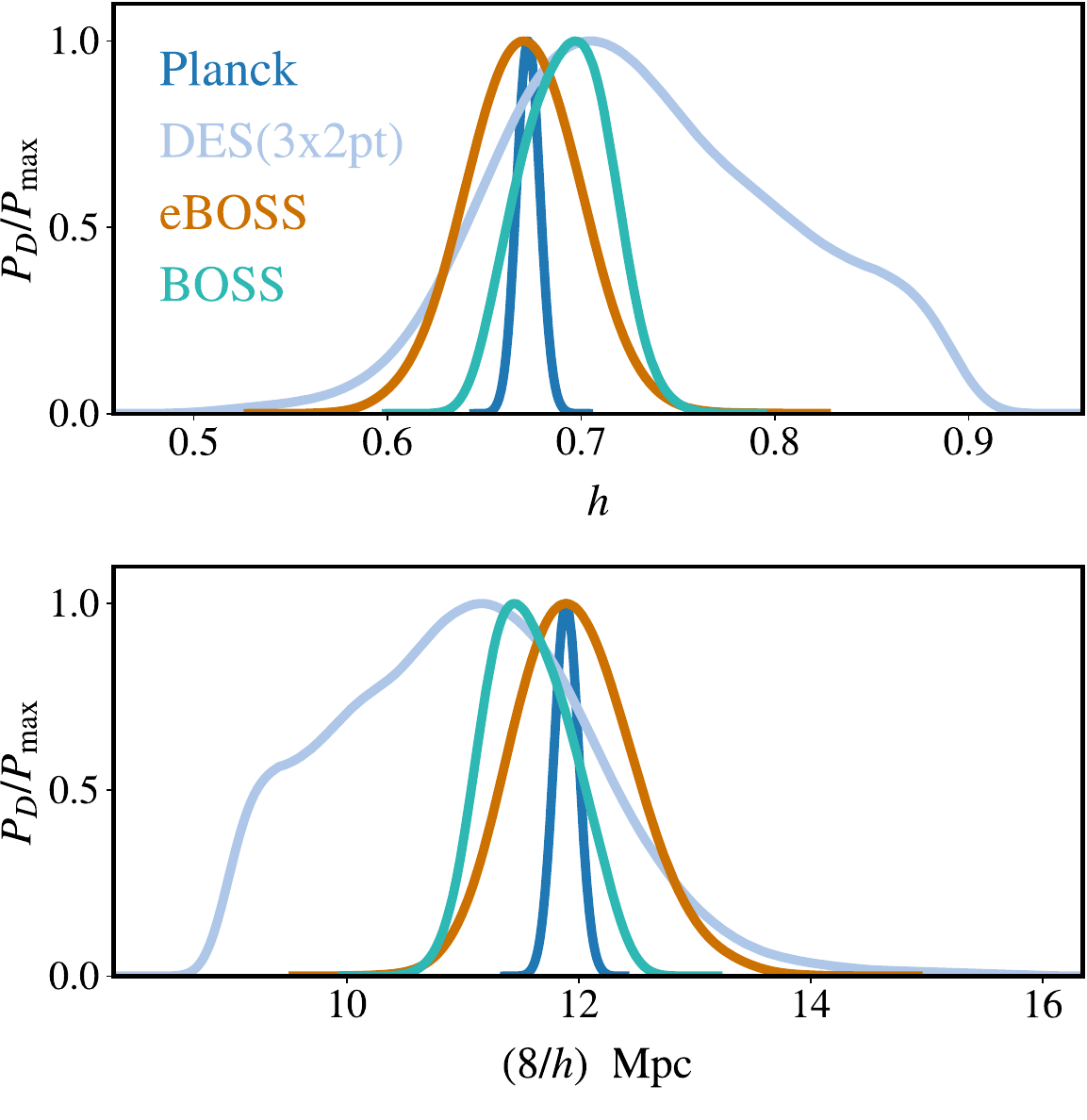}
\caption{ Upper panel\, one-dimensional marginalised posteriors for $h$ for the different data sets considered in this work (with the 
priors used in this analysis). Lower panel: the corresponding posteriors of the physical value of $(8/h)\,{\rm Mpc}$ 
- the scale used to define $\sigma_8$. Any distance defined in $h^{-1}{\rm Mpc}$ units will correspond to a range of physical scales, 
as determined by the posterior of $h$. If the posterior is prior limited, as is the case with weak lensing, the choice of prior will also 
influence the range of physical scales that contribute to $\sigma_8$. On the other hand, the effect is much smaller for the case of 
narrow Gaussian $h$-posterior - for Planck $\sigma_8$ will correspond to a  scale of $12\,{\rm Mpc}$.}
\label{fig:hpost}
\end{figure}

The main consequence of using quantities that depend on $h$ in cosmological analyses is that this 
complicates the comparison of constraints  derived from probes that lead to different posterior 
distributions on $h$. 
This can be illustrated the most straightforwardly when considering $\sigma_8$, which is 
defined in terms of a scale in $h^{-1}{\rm Mpc}$ units. As done by  \citet{sanchez2020let},  
the one-dimensional marginalised posterior distribution for $h$ can be used to obtain the 
corresponding posterior for $(8/h)\,{\rm Mpc}$ to explore what physical distances this 
radius corresponds to. Fig. \ref{fig:hpost} repeats this simple exercise 
for the data sets considered in this work - as expected, the range of scales recovered in each case heavily depends on the type of probe considered (Planck displaying an extremely narrow posterior at the physical 
scale of approximately $12\,{\rm Mpc}$, while the remaining probes cover varying ranges), especially in the 
case where the posterior of $h$ is simply limited by the prior imposed, as is the case for weak 
lensing data sets.

\begin{figure}
\centering
	\includegraphics[width=0.95\columnwidth]{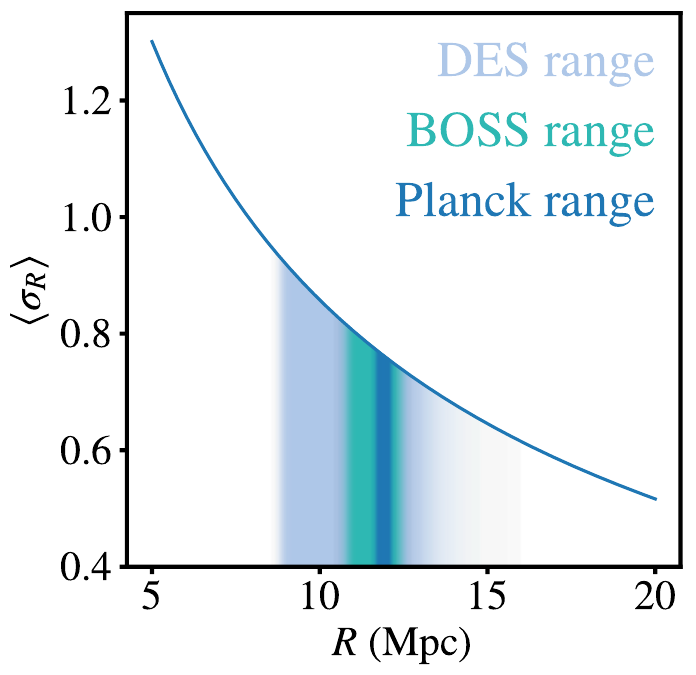}
    \caption{The change of the value of standard deviation of linear matter fluctuations $\sigma_R$ 
    measured in a sphere of physical radius $R$ in Mpc in Planck $\Lambda$CDM Universe. The shaded 
    areas indicate the ranges that $(8/h)\,{\rm Mpc}$ correspond to for BOSS, DES and Planck based on the posteriors in Fig. \ref{fig:hpost}. When $R$ is defined in $h^{-1}$Mpc, as is the case for 
    $\sigma_8$, the value measured is, in fact, a weighted average of $\sigma_R$ over a range of $R$. }
    \label{fig:sigmar}
\end{figure}

The solid line in Fig. \ref{fig:sigmar} shows the density 
field variance $\sigma_R$ as a function of the scale $R$ in a Planck $\Lambda$CDM Universe. 
The shaded areas indicate the range of physical scales covered by the the posterior distributions of 
$(8/h)\,{\rm Mpc}$ for DES, BOSS, and Planck shown in Fig.~\ref{fig:hpost}. The issue with $\sigma_8$ is then that its marginalized value corresponds to a weighted average of 
$\sigma_R$ on a range of scales that is different for each data set.
A further complication is that the value of $h$ also has an impact on the amplitude of $\sigma_R$. As discussed in \citet{sanchez2020let}, these issues can be avoided by considering the variance of 
the density field on a reference scale in Mpc such as $\sigma_{12}$, which is equivalent to $\sigma_8$ 
but is defined on a physical scale of $12\,{\rm Mpc}$. We, therefore, opt to focus on $\sigma_{12}$ 
and quantities that carry no explicit dependence on the Hubble constant $h$ in order to enable us to 
appropriately combine and compare the constraints from our data sets.

\begin{figure*}
\includegraphics[width=0.99\columnwidth]{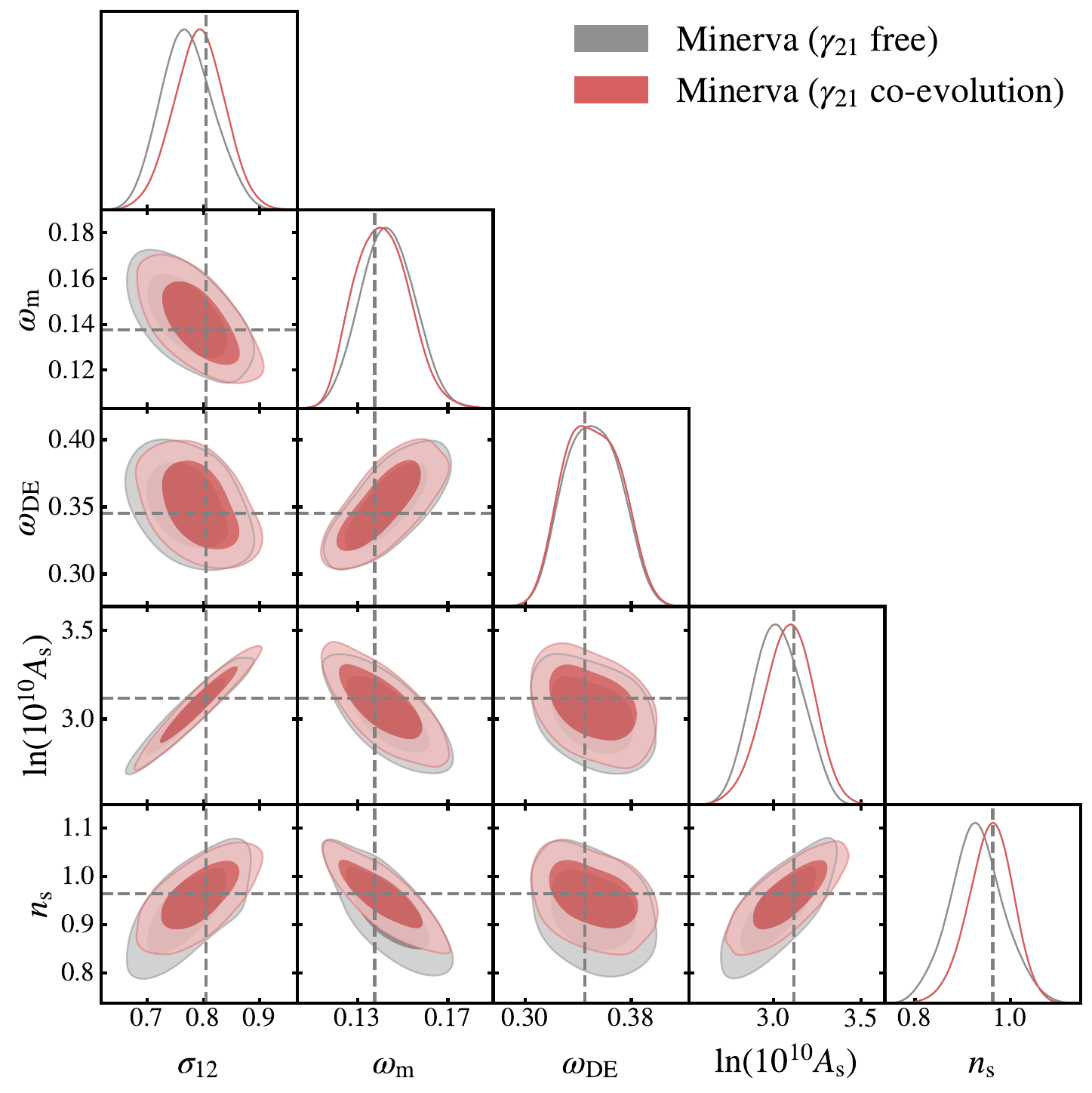}
\includegraphics[width=0.99\columnwidth]{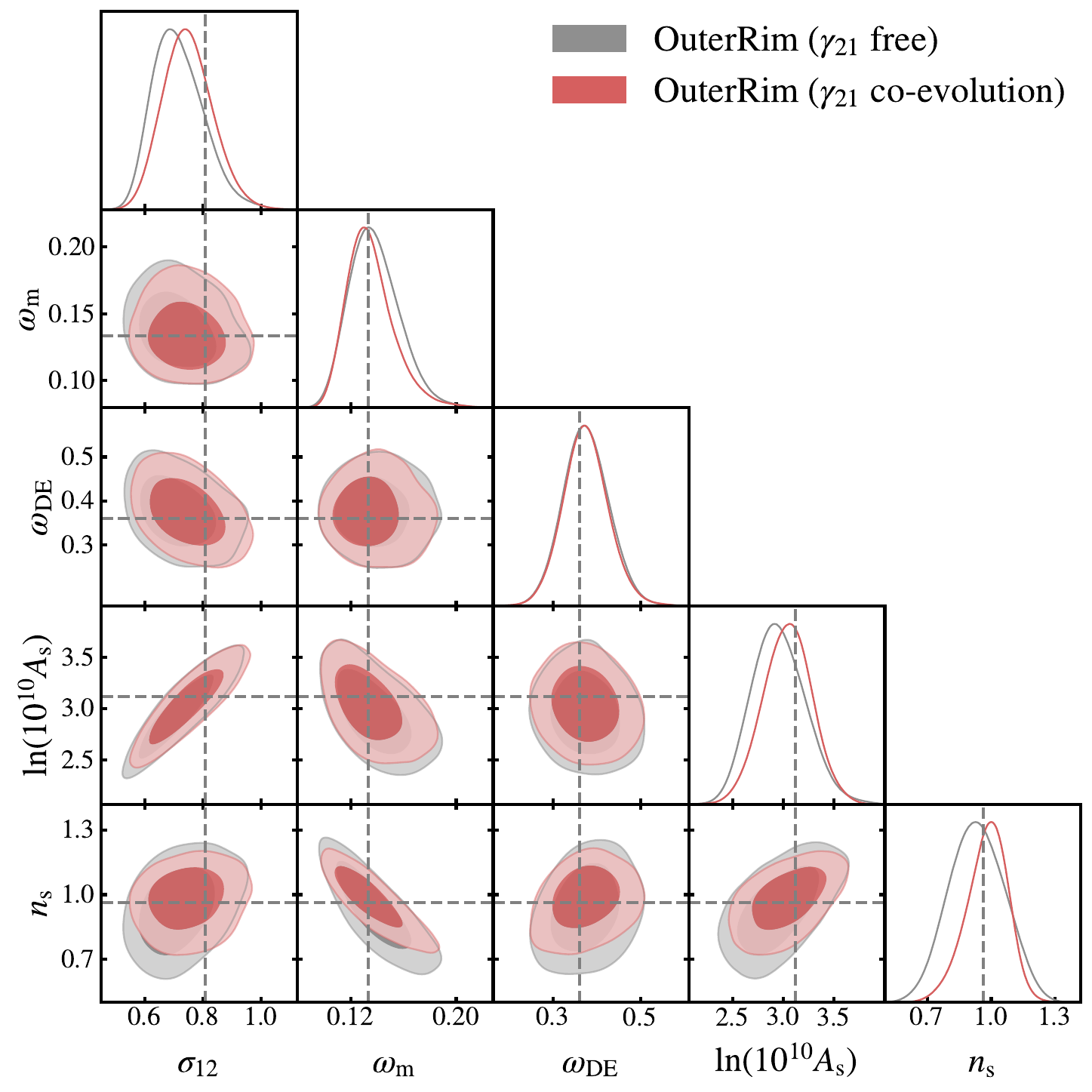}
\caption{Flat $\Lambda$CDM constraints derived from mean measurements of \textsc{Minerva} (left) 
and  \textsc{OuterRim} (right) HOD samples using the model described in Sec.~\ref{section:model} 
while freely varying the non-local bias parameters $\gamma_{21}$ (grey contours) and when 
its value is fixed using the  co-evolution relation of equation~(\ref{eq:coevol}) (red). The dashed lines 
mark the true input parameter values. Both cases recover the input cosmology well but 
the co-evolution relations yields slightly more accurate and precise constraints.}
\label{fig:or_validation}
\end{figure*}
We obtain the posterior distribution of all these parameters by performing Monte Carlo Markov chain 
(MCMC) sampling with {\sc CosmoMC} \citep{Lewis_2002}, which uses {\sc CAMB} to calculate the 
linear-theory matter power spectra \citep{Lewis_2000}, adapted to compute the theoretical model of 
our anisotropic clustering measurements described in Sec.~\ref{section:model}.
{\sc CosmoMC} uses as basis parameters the set
\begin{equation}
\pmb{\theta}_{\rm base} = \left(\omega_{\rm b},\omega_{\rm c},\Theta_{\rm MC}, A_{\rm s}, n_{\rm s} \right),
\label{eq:base_cosmomc}
\end{equation}
where $\Theta_{\rm MC}$ is defined by a factor 100 times the approximate angular size of the 
sound horizon at recombination. 
With the exception of the physical baryon density, we assign flat uninformative priors to all the 
parameters of equation~(\ref{eq:base_cosmomc}) as was done in \citet{Tr_ster_2020}. 
Our prior for $\omega_{\rm b}$ has to be restrictive, as our clustering measurements cannot 
constrain this parameter by themselves. Nevertheless, it is chosen to be approximately 
25 times wider than the constraints on this parameter derived from \textit{Planck} data alone \citep{Planck2018}.
Even though we do not sample the Hubble parameter $h$, we still need to specify the values 
allowed - our chosen range $0.5 < h< 0.9$ is wider than that of the KiDS-450 analysis of \citet{Hildebrandt_2016}
and comparable to the one used in the DES-YR1 fiducial analysis of \citet{Abbott2017b}. 
\citet{Joudaki_2016} showed that the prior on $h$ has no impact on the significance of the 
$\sigma_8$ tension. We list all the priors used in this analysis in Table \ref{tab:priors}.

As discussed by \citet{Sanchez2021}, the effect of all evolution parameters on the linear matter power spectrum is degenerate:
for a given set of shape parameters, the linear power spectra of all possible 
combinations of evolution parameters that lead to the same value of $\sigma_{12}(z)$
 are identical. This behaviour is inherited by the non-linear matter power spectrum predicted
 by {\sc respresso}, which depends exclusively on $P_{\rm L}(k)$. 
However, the full model of $P(k,\mu)$ does not follow this simple degeneracy due to the effect
of bias, RSD and AP distortions. 
Of the parameters listed in equation~(\ref{eq:base_lcdm}), $\omega_{\rm b}$, $\omega_{\rm c}$, 
and $n_{\rm s}$ are shape parameters, while $\omega_{\rm DE}$ and $A_{\rm s}$ are 
purely evolution parameters. Other quantities such as $h$ or $\Omega_i$ represent a 
mixture of both shape and evolution parameters.

Present-day CMB measurements can constrain the values of most shape parameters with high 
accuracy, with posterior distributions that are well described by a multivariate Gaussian, 
independently of the evolution parameters being explored. On the other hand, clustering 
measurements on their own provide only weak constraints on the values of the shape parameters.
However, if the shape parameters are fixed, clustering data can provide precise measurements 
of the evolution parameters. 
To test the impact of the additional information on the shape of the linear power spectrum, 
along with the priors described above, we use another set of priors to explore the constraints 
on the evolution parameters. For these runs, we impose Gaussian priors on the cosmological 
parameters that control the shape of the linear
power spectrum - $\omega_{\rm b}$, $\omega_{\rm c}$, and $n_s$. 
We derived the covariance matrix and mean values for these priors 
from our Planck-only posterior distributions. 
We refer to these constraints as the `Planck shape' case. 

\begin{table*}
	\centering
	\caption{Marginalised posterior constraints (mean values with 68 per-cent confidence interval) derived from the full 
	shape analysis of BOSS + eBOSS clustering measurements on their own, as well as in combination with the 
	$3\times 2$pt measurements from DES Y1 and the CMB data from Planck. We present two sets of constraints: our main results derived with wide priors, as listed in Table \ref{tab:priors}, and the `Planck shape' constraints obtained by imposing narrow Gaussian priors on the cosmological parameters controlling the shape of the linear power spectrum: the physical baryon density $\omega_{\rm{b}}$, the physical cold dark matter density $\omega_{\rm{c}}$ and the spectral index $\rm{n}_{\rm{s}}$, as discussed in Sect. \ref{sec:parameters}. }
	\label{tab:constraints}
	\begin{tabular}{cccccc} 
		\hline
		\multicolumn{4}{c|}{Wide priors} & \multicolumn{2}{c}{Gaussian priors on $\omega_{\rm{b}}$, $\omega_{\rm{c}}$, $\rm{n}_{\rm{s}}$}  \\
		\hline
		Parameter & BOSS~+~eBOSS & \makecell{BOSS~+~eBOSS\\+ DES} & \makecell{BOSS + eBOSS\\+ DES + Planck} & BOSS~+~eBOSS& \makecell{BOSS~+~eBOSS\\+ DES}\\
		\hline
		$\sigma_{12}$ & 0.805 $\pm$ 0.049 & $0.795^{+0.032}_{-0.037}$ & 0.7890 $\pm$ 0.0078  & 0.785 $\pm$ 0.039 & 0.766 $\pm$ 0.019\\
		$\omega_{\rm{m}}$ & 0.134 $\pm$ 0.011 & 0.131 $\pm$ 0.011 & 0.14090$\pm$ 0.00085 & 0.1426 $\pm$ 0.0013 & 0.1423 $\pm$ 0.0012\\
		$\omega_{\rm{DE}}$ & 0.328 $\pm$ 0.020 & 0.327 $\pm$ 0.020 & 0.3268 $\pm$ 0.0064  & $0.327^{+0.011}_{-0.013}$ & 0.335 $\pm$ 0.011\\
		$\rm{ln}10^{10}\rm{A}_{\rm{s}}$ & 3.13 $\pm$ 0.15 & 3.14 $\pm$ 0.13 & 3.041 $\pm$ 0.016 & 3.011 $\pm$ 0.099 & 2.976 $\pm$ 0.054\\
		$n_{\rm s}$ & 1.009 $\pm$ 0.048 & 1.001 $\pm$ 0.047 & 0.9700 $\pm$ 0.0038 & 0.9660 $\pm$ 0.0044 & 0.9665 $\pm$ 0.0043\\
		\hline
		$\sigma_8$ & 0.815 $\pm$ 0.044 & 0.803 $\pm$ 0.028 & 0.8029 $\pm$ 0.0066 & 0.800 $\pm$ 0.039 & 0.785 $\pm$ 0.021 \\
		$\Omega_{\rm{m}}$ & $0.290^{+0.012}_{-0.014}$ & $0.286^{+0.011}_{-0.013}$ & 0.3014 $\pm$ 0.0053 & 0.3037 $\pm$ 0.0081  & 0.2985 $\pm$ 0.0072\\
		$h$ & $0.679 \pm 0.021$ & $0.677 \pm 0.021$ & $0.6838 \pm 0.0041$ & $0.6855^{+0.0084}_{-0.0094}$ & $0.6905\pm 0.0083$ \\ 
		$S_8$ & 0.801 $\pm$ 0.043 & 0.783 $\pm$ 0.020 & 0.805 $\pm$ 0.011 & 0.805 $\pm$ 0.042 & 0.783 $\pm$ 0.019\\
		\hline
	\end{tabular}
\end{table*}

\subsection{Model validation}
\label{section:validation}

As we are using an updated prescription for the modelling of both the non-linear matter power 
spectrum and galaxy bias compared to the previous work of 
\citet{Tr_ster_2020}, we want to assess if it can recover unbiased cosmological parameter estimates, using mock data 
based on numerical simulations as a testing ground. We do so by applying our model to 
the mocks that were used for model validation in the original analyses: the \textsc{Minerva}
simulations \citep{Grieb2016,Lippich2019} for a BOSS-like sample and 
\textsc{OuterRim} \citep{Heitmann_2019} for an eBOSS-like data set. 

\textsc{Minerva} mocks are produced from a set of 300 N-body simulations with  $1000^3$ particles 
and  a box size of $L=1.5\,h^{-1}{\rm Gpc}$. The snapshots 
at $z=0.31$ and $z=0.57$ were used to create halo catalogues with a minimum halo mass 
of $M_{\rm{min}}=2.67\times 10^{12}h^{-1}\rm{M}_\odot$, 
which were populated with synthetic galaxies using the halo occupation distribution (HOD)
model by \citet{Zheng_2007} with  parameters designed to reproduce the clustering 
properties of the LOWZ and CMASS galaxy samples from BOSS. 

The \textsc{OuterRim} \citep{Heitmann_2019} simulation uses $10\,240^3$ dark matter 
particles to trace the dark matter density field in a $L=3\,h^{-1}{\rm Gpc}$ size box. 
We use a set of 100 mock catalogues constructed from the snapshot at $z=1.433$, which 
was populated using an HOD model matching the clustering of the eBOSS QSO sample and 
tested extensively in the mock challenge \citep[labeled as HOD0 in][]{Smith_mocks}. 
These realizations include catastrophic redshift failures at a rate of 1.5\%, which 
corresponds to that of the eBOSS quasars.  

We measured the mean clustering wedges of the samples from \textsc{Minerva} and the
Legendre multipoles from \textsc{OuterRim} with the same binning and range of scales as those of 
the real data from BOSS and eBOSS and computed their corresponding theoretical covariance matrices 
using the Gaussian recipe of \citet{Grieb2016}.  We analysed these measurements using identical 
nuisance and cosmological parameter priors as for our final results and tested the validity 
of the model described in Sec.~\ref{section:model} with and without the assumption of the 
co-evolution relation for $\gamma_{21}$ of equation~(\ref{eq:coevol}). We performed a 
joint fit of the two BOSS-like samples from \textsc{Minerva} while the \textsc{OuterRim} measurements, 
which correspond to a different cosmology, were analysed separately. Fig. \ref{fig:or_validation} 
shows the posterior distributions recovered from these measurements, which are in excellent 
agreement with the true input cosmology for all cases (shown by the dashed lines). 
Nevertheless, we find that setting the value of $\gamma_{21}$ according to 
equation (\ref{eq:coevol}) recovers the true parameter values more accurately for both 
samples and results in tighter constraints than when it is freely varied. We therefore adopt this 
approach in the analysis of the clustering measurements from BOSS and eBOSS.

\begin{figure*}
\includegraphics[width=0.99\columnwidth]{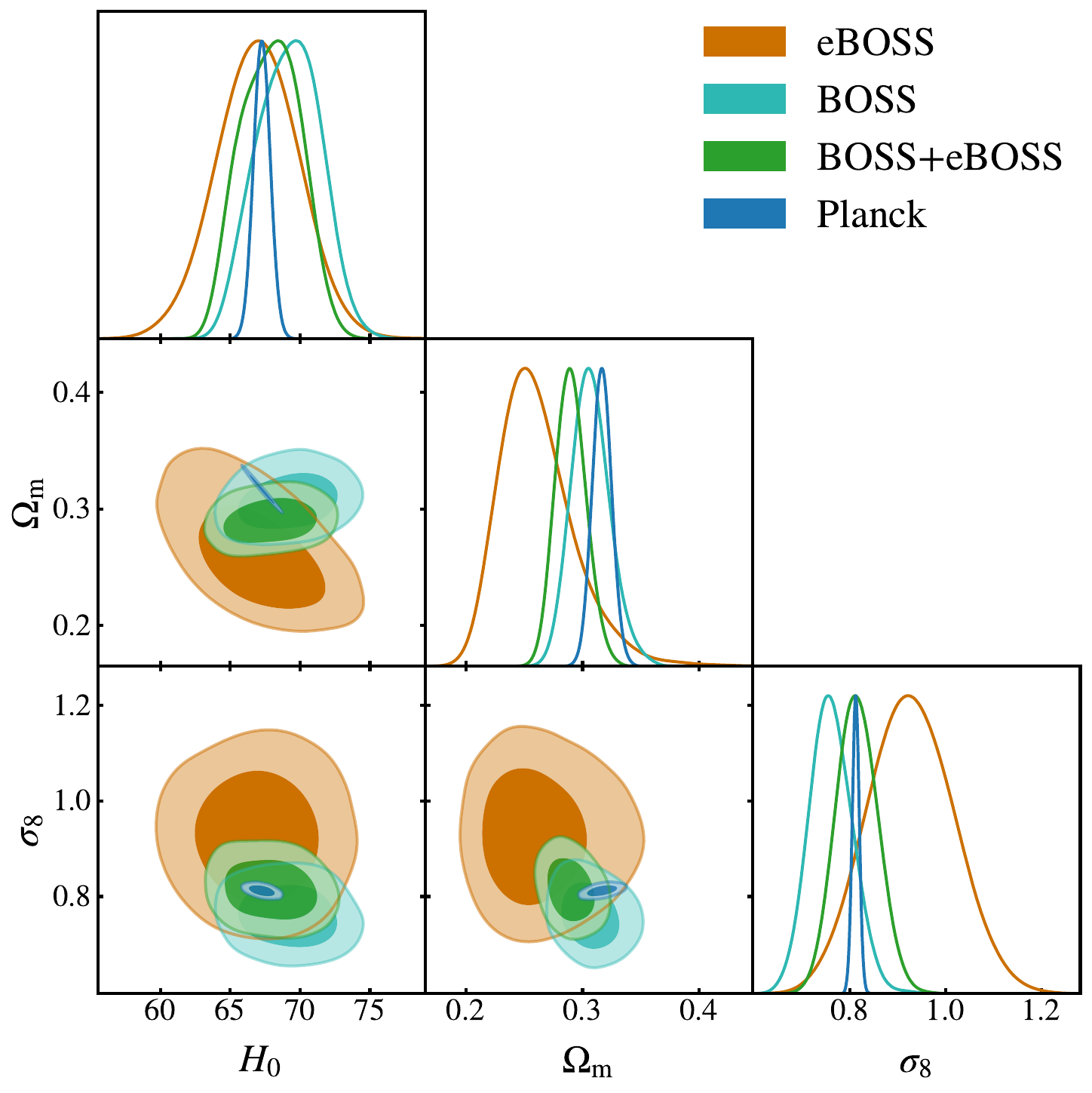}
\includegraphics[width=0.99\columnwidth]{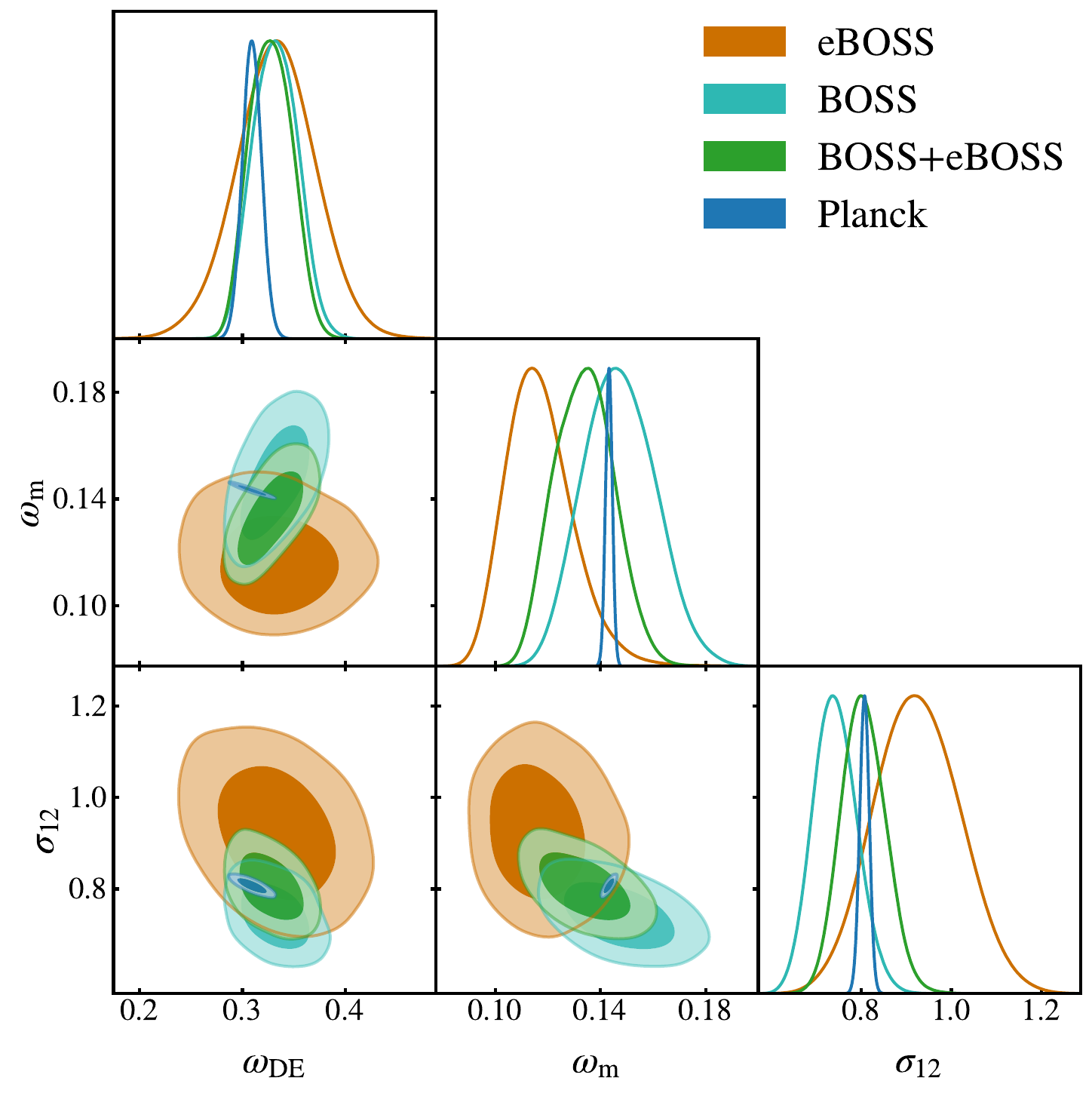}
\caption{Marginalised posterior contours in the `traditional' and $h$-independent parameter spaces 
from  the Legendre multipoles of eBOSS QSO sample (orange) and the clustering wedges of 
BOSS DR12 galaxies (light blue) for a flat $\Lambda$CDM model. The joint constraints are shown in green, 
with Planck in dark blue for comparison.   }
\label{fig:boss+eboss_s8}
\end{figure*}

\section{Results}
\label{section:results}

Our main results come from the combination of the full shape analyses of the 
BOSS galaxy clustering wedges and eBOSS QSO Legendre multipoles described in Sec.~\ref{sec:data}. 
We also present combined late-Universe constraints obtained from the joint analysis of these clustering 
measurements with the $3\times 2$pt  data set from DES Y1. 
For comparison, in Appendix~\ref{appendix:kids} we present the constraints obtained using instead the 
cosmic shear measurements from KiDS-450, which lead to similar results. 
As we find a good agreement between BOSS~+~eBOSS~+~DES and Planck, we also present
the parameter constraints obtained from the combination of all four data sets. 
These constraints are summarized in Table \ref{tab:constraints} and are discussed in 
Sects. ~\ref{sec:clustering} -- \ref{sec:joint}.

\subsection{Clustering constraints}
\label{sec:clustering}

Here we present the main result of our work - the combined flat $\Lambda$CDM constraints 
from the anisotropic clustering measurements from BOSS and eBOSS. 
Fig.~\ref{fig:boss+eboss_s8} shows the posterior distributions for BOSS and eBOSS separately 
(light blue and orange contours, respectively) as well as their combined constraints (green contours) for 
two sub-sets of cosmological parameters. For comparison, we also show the Planck-only 
constraints in dark blue. The panels on the left show the results on the more traditional 
parameter set of $\sigma_8$, $\Omega_{\rm{m}}$, and $H_0$ whereas the ones on the right 
correspond to the alternative basis discussed in Sec.~\ref{sec:parameters} of $\sigma_{12}$, 
$\omega_{\rm{m}}$, and $\omega_{\rm{DE}}$. 

Regardless of the parameter space considered, we find all of our data sets to be in good 
agreement with each other. The largest deviation between the joint BOSS~+~eBOSS constraints 
and those recovered from Planck can be observed in the matter density $\Omega_{\rm{m}}$, which 
displays a difference at the 1.7$\sigma$ level. Nevertheless, this deviation does not indicate a similarly significant 
disagreement in the physical matter density preferred by these probes, as the 
value of $\omega_{\rm m}$ recovered by our clustering constraints matches that 
of Planck within $0.8\sigma$. 
This suggests that the differences seen in $\Omega_{\rm{m}}$ are related to the 
posterior distributions on $h$ recovered from these data sets. 
Indeed, looking at our $h$-independent parameter space, we see 
that the marginalised constraint of the physical dark energy density also differs from the 
value preferred by Planck by $0.8\sigma$, 
with clustering measurements preferring slightly higher $\omega_{\rm{DE}}$, which translates into a 
higher value for $H_0$ and a lower $\Omega_{\rm{m}}$. 

\citealp{Tr_ster_2020} found that the clustering measurements from BOSS wedges prefer a 
2.1$\sigma$ lower value of $\sigma_8$ as compared to Planck. Here we confirm the low 
preference, albeit with much lower significance due to the differences in the modelling of the 
power spectrum, for both $\sigma_8$ and $\sigma_{12}$ 
(consistent with Planck at the 1.1$\sigma$ and 1.3$\sigma$ level, respectively). The increased 
consistency between these results is mainly due to the tighter constraints enabled by the use 
of the co-evolution relation of equation~(\ref{eq:coevol}), which restricts the 
allowed region of the parameter space to higher values of $\sigma_8$ and $\sigma_{12}$, as can be seen in Fig. \ref{fig:or_validation}. 
The constraints on $\sigma_8$ and $\sigma_{12}$ recovered from eBOSS are at similar levels 
of agreement with Planck, however, the values recovered are 1.3$\sigma$ and 1.2$\sigma$ 
\textit{higher} than the CMB results. This is also consistent with the most recent analysis by 
\citet{Hou2021} and \citet{Neveux_ebossqso}, who found the inferred growth rate $f\sigma_8$ to be $\sim$2$\sigma$ higher than the 
$\Lambda$CDM model with the best-fitting Planck parameters. The combination of the 
clustering measurements from BOSS and eBOSS is, therefore, in an overall excellent agreement 
with Planck - with differences at the level of 0.05$\sigma$ for $\sigma_{8}$ and 
0.04$\sigma$ for $\sigma_{12}$. 
  
As discussed in Sec.~\ref{sec:parameters}, the shape parameters 
$\omega_{\rm{b}}$, $\omega_{\rm{c}}$, and $n_{\rm s}$ are all tightly constrained by \textit{Planck}
with posterior distributions that are in complete agreement with those inferred from the other 
cosmological probes considered here. 
We can, therefore, study the improvement in the constraints on the evolution parameters 
$\omega_{\rm DE}$ and $A_{\rm s}$ that are obtained from the LSS probes when 
the shape of the power spectrum is constrained to match that of Planck's cosmology. As described in Sec.~\ref{sec:parameters}, we achieve this by adding an informative Gaussian prior on the shape parameters 
based on our Planck runs and repeating our analysis with an otherwise identical set up. 

\begin{figure}
\includegraphics[width=0.99\columnwidth]{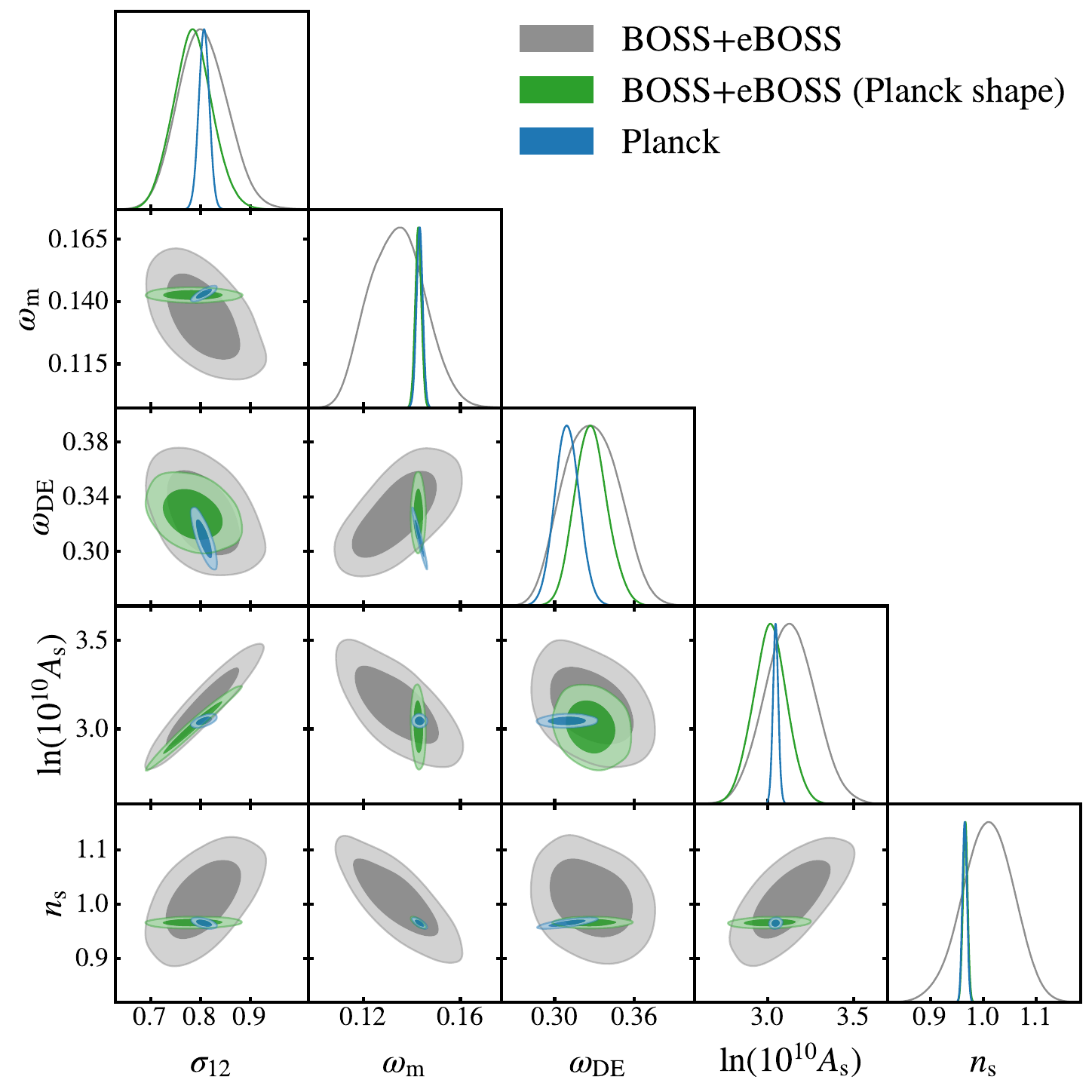}
\caption{When an informative prior is imposed on BOSS~+~eBOSS for the shape parameters 
$\omega_{\rm b}$, $\omega_{\rm c}$, and  $n_{\rm s}$ so as to match the power spectrum 
shape obtained by Planck, the recovered constraints (in green) on the evolution parameters 
are in a good agreement with Planck (dark blue) with a slightly more significant deviation in 
$\omega_{\rm{DE}}$ only: BOSS~+~eBOSS prefer a $\sim$1.2$\sigma$ higher value of 
$\omega_{\rm{DE}}$ than Planck.   }
\label{fig:boss+eboss_planckshape}
\end{figure}

The results of this exercise are shown in Fig. \ref{fig:boss+eboss_planckshape}. As 
the two data sets were already in a good agreement across the parameter space, including the shape 
parameters, imposing additional priors simply adds constraining power on the degenerate 
evolution parameters, most notably $\omega_{\rm{DE}}$ (degenerate with $\omega_{\rm{m}}$), 
which is recovered to be slightly higher than the Planck value to compensate the slight shifts in 
$\sigma_{12}$ and $\text{ln}(10^{10}A_{\rm s})$ to lower values.

\subsection{Consistency with Planck}

When looking at marginalised posteriors we are limited by our selection of the parameter space as 
well as the associated projection effects and, while we can use the standard deviation to quantify 
agreement on a particular parameter value, this becomes inappropriate when larger parameter 
spaces are considered. We, therefore, wish to further explicitly quantify the agreement between 
eBOSS~+~BOSS and Planck using a tension metric, as has become standard in cosmological analyses. 

First, we want to establish agreement over the whole parameter space considered. In order to do this, 
we use the suspiciousness tension metric, $S$,  introduced by \citealp{PhysRevD.100.043504}. The 
main advantages of using suspiciousness include the fact that it measures the agreement between 
two data sets across the entire parameter space, similarly to the Bayes factor $R$. However, unlike 
$R$, the suspiciousness is by construction insensitive to prior widths, as long as the posterior is not 
prior-limited. Given two data sets, A and B, the suspiciousness quantifies the mismatch between 
them by comparing the relative gain in confidence in data set A when data set B is added (as 
measured by $R$) with the unlikeliness of the two data sets ever matching as measured by 
the information ratio $I$, that is
\begin{equation}
    \ln S=\ln R-\ln I.
    \label{eq:def_lnS}
\end{equation}
Following the method described in \citet{kids1000}, we redefine $\ln R$ and $\ln I$ 
in terms of the expectation values of the log-likelihoods $\left<\ln\mathcal{L}\right>$ and evidences 
$Z$. The evidences, however, cancel out and we are able to calculate $S$ from the expectation values only:
\begin{equation}
    \ln S =\left<\ln\mathcal{L}_{\rm A+B}\right>_{P_{\rm A+B}}-\left< \ln\mathcal{L}_{\rm A}\right>_{P_{\rm A}}-\left<\ln\mathcal{L}_{\rm B}\right>_{P_{\rm B}}.
\end{equation}
The value of $S$ can then be interpreted using the fact that, for Gaussian posteriors, the 
difference $d-2\ln S$, where $d$ is the Bayesian model dimensionality, is $\chi^2_d$ 
distributed. We calculate $d$ for each of the data sets separately, $d_{\rm A}$ and $d_{\rm B}$, and their combination, 
$d_{\rm A+B}$, as described in \citet{PhysRevD.100.043504} and combine the results as
$d=d_{\rm A}+d_{\rm B}-d_{\rm A+B}$.

Applying this procedure to eBOSS~+~BOSS and Planck, we find $\ln S=0.41\pm 0.07$ with a Bayesian dimensionality 
of d=$4.5\pm 0.4$, which correctly indicates that there are approximately 5 cosmological parameters shared 
between the two data sets. This can then be related to a p-value of $p = 0.52 \pm 0.02$
or a tension of $0.64 \pm 0.03\sigma$, which is consistent with the $0.76 \pm 0.05\sigma$ 
tension between Planck and BOSS alone found by \citet{Tr_ster_2020}  and 
indicates a good agreement between these data sets. 

In addition to the suspiciousness, we want to use a tension metric that allows for a greater 
control to focus only on a selected subset of parameters. For this purpose, we use the update 
difference-in-mean statistic, $\mathcal{Q}_{\rm{UDM}}$, as described in \citet{raverihu19} and 
implemented in \textsc{tensiometer}\footnote{https://github.com/mraveri/tensiometer} 
\citep{lemos2020assessing}.
This statistic extends the simple difference in means, where the difference in mean parameter values $\hat{\pmb{\theta}}$ 
measured by two data sets is weighted by their covariance $\mathbfss{C}$. The `update' in UDM refers to the 
fact that instead of comparing data set A with data set B, we consider the updated information
in the combination ${\rm A+B}$ with respect to A by means of 
\begin{equation}
    \mathcal{Q}_{\rm{UDM}} = \left(\hat{\pmb{\theta}}_{\rm A+B}-\hat{\pmb{\theta}}_{\rm A}\right)^t\left(\mathbfss{C}_{\rm A}-\mathbfss{C}_{\rm A+B}\right)^{-1} \left(\hat{\pmb{\theta}}_{\rm A+B}-\hat{\pmb{\theta}}_{\rm A}\right).
\end{equation}
This has the advantage of the posterior of ${\rm A+B}$ being more Gaussian than that of B alone. 
For Gausian distributed parameters, $\mathcal{Q}_{\rm{UDM}}$ is chi-square distributed with 
a number of degrees of freedom given by the rank of 
$\left(\mathbfss{C}_{\rm A}-\mathbfss{C}_{\rm A+B}\right)$. The calculation of 
$\mathcal{Q}_{\rm{UDM}}$ may be performed by finding the Karhunen–Loéve (KL) modes 
of the covariances and re-expressing the cosmological parameters in this basis. This 
transformation allows us to reduce the sampling noise by imposing a limit to the eigenvalues 
of the modes that are considered and in this way cutting out those that are dominated by 
noise (which represent the directions in which adding B does not improve 
the constraints with respect to A). The number of remaining modes correspond to
the degrees of freedom with which $\mathcal{Q}_{\rm{UDM}}$ is distributed. For our tension 
calculations we, therefore, only select the modes $\alpha$ whose eigenvalues $\lambda_{\alpha}$ satisfy :
\begin{equation}
    0.05 < \lambda_{\alpha}-1<100.
\end{equation}
This corresponds to requiring that a mode of the base data set is updated by at least 5 per-cent. 
We subsequently find that there are 2 modes being constrained when Planck is updated by both probe combinations 
considered in this work (BOSS~+~eBOSS and BOSS~+~eBOSS~+~DES).

For BOSS~+~eBOSS we get $\mathcal{Q}_{\rm{UDM}} = 2.0$ for the full parameter space, resulting 
in a `tension' with Planck of $0.90\sigma$ - only slightly higher than what $S$ suggests.

\begin{figure}
\includegraphics[width=0.99\columnwidth]{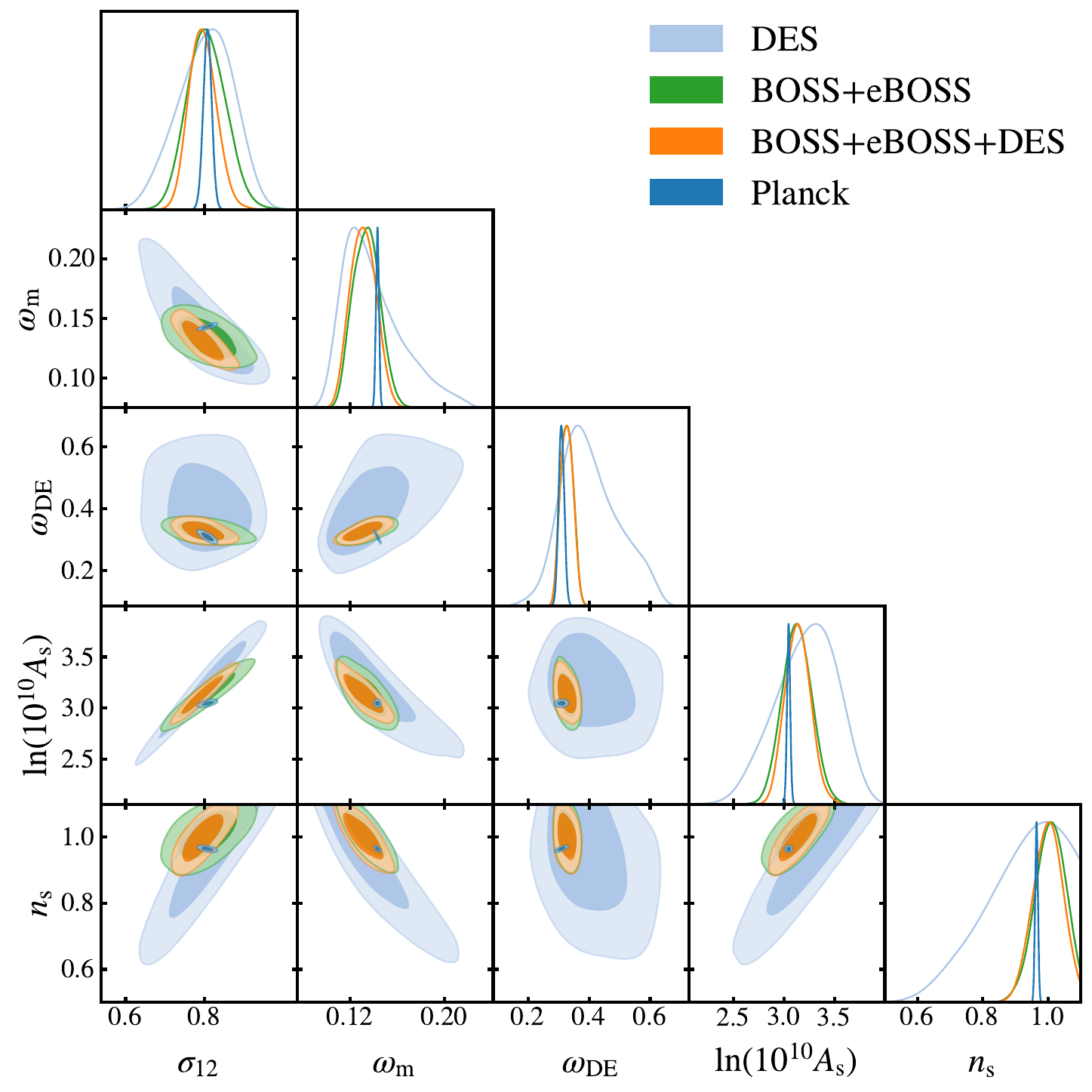}
\caption{In orange - `low-redshift' constraints for flat $\Lambda$CDM  obtained from 
combining BOSS~+~eBOSS clustering (green) with DES $3\times 2$pt (light blue) and compared 
with Planck (dark blue). While we obtain a good consistency overall, we note the 
slight discrepancy between the low redshift probes and Planck contours in 
$\log(10^{10}A_{\rm{s}})-\sigma_{12}$ and $\omega_{\rm m}$--$\sigma_{12}$ 
projections, reminiscent of the tension seen in $\sigma_8$--$\Omega_{\rm m}$ plane. }
\label{fig:lowz}
\end{figure}

\subsection{Joint analysis with DES data}
\label{sec:joint}

Following \citet{Tr_ster_2020}, we want to further investigate the constraints from multiple 
low-redshift probes together by adding a weak lensing data set - in this case, the 
$3\times2$pt measurements from DES Y1.  \citet{Tr_ster_2020} used the suspiciousness statistic
and showed that the combination of BOSS clustering and KiDS-450 shear measurements are 
in $\sim$~$2\sigma$ tension with Planck. The most recent KiDS-1000 $3\times2$pt analysis 
\citep{kids1000}, where the BOSS galaxy sample was used for galaxy clustering and 
galaxy-galaxy lensing measurements, also found a similar level of tension when the entire 
parameter space is considered. As DES Y1 measurements have no overlap with either 
BOSS or eBOSS, we can treat these data sets as independent and easily combine them to 
test whether we also find a similar trend.

The resulting constraints are presented in Fig.~\ref{fig:lowz}. We confirm that 
DES is in good agreement with eBOSS~+~BOSS (with $\ln S=-1.08\pm 0.05$, which 
corresponds to a $1.3\pm 0.08\sigma$ tension) and it is, therefore, safe to combine 
them. The addition of DES data to the analysis provides only slightly tighter constraints
with respect to eBOSS~+~BOSS, with the greatest improvement in $\sigma_{12}$, and 
an overall good agreement with Planck.

\begin{figure}
\includegraphics[width=0.99\columnwidth]{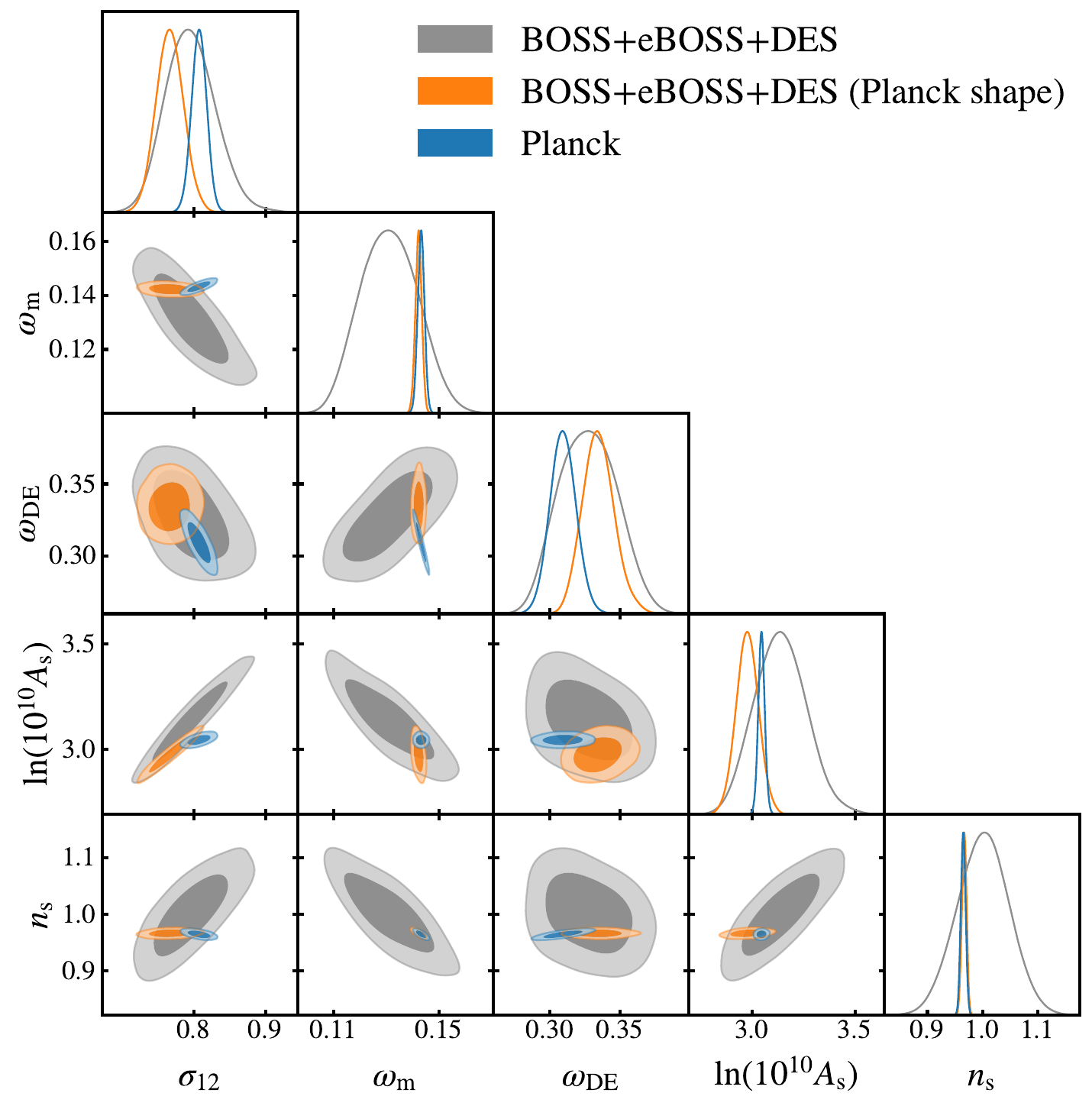}
\caption{
Constraints on flat $\Lambda$CDM models from the full combination of low-redshift 
probes (DES~+~BOSS~+~eBOSS) obtained after imposing a Planck prior on shape parameters 
$n_{\rm s}, \omega_{\rm b}$ and $\omega_{\rm c}$ (orange contours). The constraints 
from the original uninformative prior analysis (grey contours) and Planck (dark blue contours) are shown for 
comparison. The results show similar trends as in the case of BOSS~+~eBOSS. 
There is a shift to higher values of $\omega_{\rm DE}$ that leads to a lower power 
spectrum amplitude today, $\sigma_{12}$, and, to a lesser extent, a lower 
$\log(10^{10} A_{\rm s})$.}
\label{fig:lowz_shape}
\end{figure}

Nevertheless, it is worth noting that, when considering the two dimensional 
posterior projections, there are two parameter combinations in particular for 
which the $1\sigma$ contours of DES~+~BOSS~+~eBOSS and Planck do not overlap. 
The slight discrepancy we observe in the $\omega_{\rm{m}}$ -- $\sigma_{12}$ plane 
is reminiscent of the `$\sigma_8$ tension' seen in $\Omega_{\rm{m}}$ -- $\sigma_{8}$ 
and is larger than the discrepancy displayed by either of the probes individually. 
In addition to that, we also see a similarly slight disagreement in the 
$\ln(10^{10}A_{\rm s})$ -- $\sigma_{12}$ plane. 
The projection of $A_{\rm s}$ 
with $\sigma_{12}$ (as opposed to $\sigma_8$) allows us to recover the tight degeneracy 
between the two parameters which exposes how, for a given present-day clustering amplitude, 
low-redshift probes prefer a higher initial power spectrum amplitude. 

We find that adding the DES Y1 $3\times 2$pt measurements worsens the agreement 
with Planck with respect to the results obtained from the combination of BOSS and eBOSS alone.
We obtain a suspiciousness of $\ln S=-1.86 \pm 0.04$, corresponding to a tension of $1.54 \pm 0.08\sigma$. 
When considering the UDM statistic across the entire shared parameter space, we 
find $\mathcal{Q}_{\rm{UDM}}=6.3$ distributed with 2 degrees of freedom, which translates 
into a tension at the $1.9\sigma$ level. As for the case of the clustering-only constraints, 
$\mathcal{Q}_{\rm{UDM}}$ indicates a greater level of tension than $S$.

\citet{lemos2020assessing} found that the DES Y1 $3\times 2$pt measurements alone 
are in a 2.3$\sigma$ tension with Planck, as measured by $\mathcal{Q}_{\rm{UDM}}$. This increases 
to $2.4  \pm  0.02\sigma$ when using the suspiciousness statistic. These levels of tension are 
comparable with what we find from the full combination of low-redshift probes. The tension between 
Planck and weak lensing data sets is usually interpreted as a reflection of tension in the parameter 
combination $S_8 = \sigma_8(\Omega_{\rm{m}}/0.3)^{0.5}$, that is taken to describe the 
`lensing strength'. The $S_8$ value that we recover from the joint low redshift probes is also about 2$\sigma$ lower than the Planck constraint (see Table \ref{tab:constraints}). Nevertheless, as we see in Fig. \ref{fig:lowz}, there is a comparable discrepancy in $\log(10^{10}A_{\rm{s}})$ -- $\sigma_{12}$ plane.
We can use $\mathcal{Q}_{\rm{UDM}}$ in order to quantify and compare the level of tension present in 
these two-dimensional projections by calculating it for a subset of shared parameter space. We find that the 
amount of tension in both $\Omega_{\rm{m}}$ -- $\sigma_{8}$ and its $h$-independent equivalent 
is $\sim2.0\sigma$, whereas $\log(10^{10}A_{\rm{s}})-\sigma_{12}$ displays a slightly higher tension 
of 2.5$\sigma$. 

We also repeated our fitting procedure with an additional Gaussian prior on the parameters controlling 
the shape of the power spectrum to be consistent with Planck, as described in Sec.~\ref{sec:parameters}. 
The resulting posteriors are shown in Fig.~\ref{fig:lowz_shape}. We observe the same general trends as 
from the analysis of our clustering data alone discussed in Sec.~\ref{sec:clustering}. However, 
the prior on the shape parameters leads to larger shifts in the evolution parameters. 
This is expected, as DES data on their own cannot constrain the shape parameters well. Adding the 
informative priors breaks the denegeracies between shape and evolution parameters and increases 
the constraining power significantly. This, 
in turn, exposes  any discrepancies in the evolution parameters. The values of 
$\sigma_{12}$ and $\ln(10^{10}A_{\rm s})$ preferred by our low-redshift probes 
when an informative prior is imposed are, respectively, $1.89\sigma$ and $1.22\sigma$ lower 
than the corresponding Planck values. Meanwhile, the recovered value for 
$\omega_{\rm{DE}}$ is  $1.73\sigma$ higher.

\section{Discussion}
\label{section:discussion}

The flat $\Lambda$CDM constraints from the low-redshift probes presented in Sec.~\ref{section:results} 
show a consistent picture. Updating the power spectrum model and 
supplementing the clustering measurements with eBOSS data brings the joint BOSS~+~eBOSS constraints 
to a better agreement with Planck than the BOSS-only results from \citet{Tr_ster_2020}. 
These constraints are not significantly modified when these data are combined with DES, 
resulting in a good overall consistency with Plank across the entire parameter space, as indicated by 
both $S$ and $\mathcal{Q}_{\rm{UDM}}$.

Nevertheless, when considering specific two-dimensional projections we still see intriguing differences, 
mainly driven by the lensing data.
Although the constraints in the $\sigma_{12}$ -- $\omega_{\rm{m}}$ plane obtained using BOSS~+~eBOSS
and DES data separately do not show the discrepancy with Planck that characterizes the results in their 
$h$-dependent counterparts of $\sigma_8$ and $\Omega_{\rm{m}}$, 
the full combination of low-redshift probes tightens the degeneracy between these parameters 
and leads to constraints that are just outside the region of the parameter space preferred by Planck.

We also see differences in the $\log(10^{10}A_{\rm s})$ -- $\sigma_{12}$ plane between DES and 
Planck, which are inherited by the full combination of low-redshift data sets. 
The tight relation between these parameters, which is not seen when using $\sigma_8$, illustrates 
the closer link between $\sigma_{12}$ and the overall amplitude of density fluctuations obtained
by eliminating the ambiguity caused by the dependency on $h$.  
For a given value of $\sigma_{12}$, Planck measurements prefer a lower initial amplitude of density 
fluctuations than DES, suggesting a discrepancy in the total growth of structures predicted by these
two data sets. 

Within the context of a $\Lambda$CDM model, the key parameter controlling the 
growth of structure at low redshift is the physical dark energy density. Indeed, as 
can be seen in Fig.~\ref{fig:lowz} the posterior distribution of $\omega_{\rm DE}$ recovered
from DES extends to significantly higher values than the one obtained using Planck CMB measurements. 
The tendency of the low-redshift data to prefer a higher value of $\omega_{\rm DE}$ than 
that of Planck can be seen more clearly in the results obtained after imposing a prior on the
shape parameters shown in Fig.~\ref{fig:lowz_shape}. In this case, we find $\omega_{\rm DE} = 0.335 \pm 0.011$
using BOSS~+~eBOSS~+~DES while Planck data lead to $\omega_{\rm DE} = 0.3093 \pm 0.0093$. 
A higher value of $\omega_{\rm DE}$ corresponds also to a higher value of $h$. 
Therefore, this difference is also interesting in the context of the Hubble tension, as many of 
the proposed solutions to this issue focus on modifying the dark energy component. 

\begin{figure}
\centering
	\includegraphics[width=1.0\columnwidth]{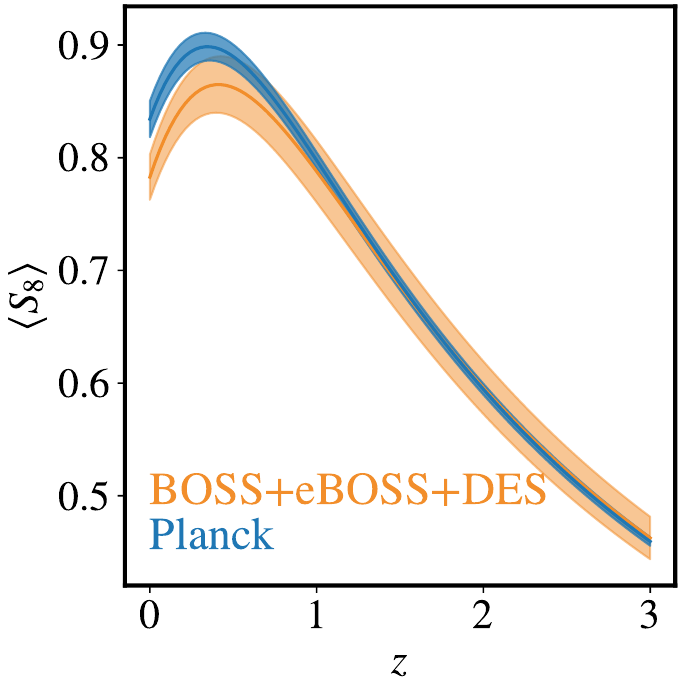}
    \caption{Comparison of the inferred mean value for $S_8(z)$ (solid lines) and their corresponding 68 
    per-cent confidence level (shaded area) corresponding to the combination of BOSS~+~eBOSS~+~DES 
    (orange) and Planck (blue). }
    \label{fig:S8_z}
\end{figure}

The analysis of the consistency between low- and high-redshift data has been focused on 
the comparison of constraints on $S_8$,  which depends on the present-day value of $\sigma_8$. 
Fig.~\ref{fig:S8_z} shows the redshift evolution of $S_8(z)$ predicted by Planck and 
the combination of all low-redshift data sets. These curves are consistent at 
high redshift during matter domination and start to diverge at $z < 1$ to reach a difference  
at the 2$\sigma$ level at $z = 0$. 
However, as this redshift is not probed by any LSS data set, the value of $S_8(z=0)$
is an extrapolation based on the assumption of a $\Lambda$CDM background evolution. 
Extending this extrapolation to $a > 1$, the difference between the two cosmologies  
continues to increase and becomes even more significant. 
Therefore, quoting the statistical significance of any discrepancy in the recovered 
values of $S_8(z=0)$ might not be the best characterization of the difference in the 
cosmological information content of these measurements. 

As discussed before, DES and Planck data appear to prefer different evolutions for the 
growth of cosmic structure, which in a $\Lambda$CDM universe depends on 
$\omega_{\rm m}$ and $\omega_{\rm DE}$. As the former is exquisitely constrained by 
Planck for general parameter spaces, the latter is perhaps the most interesting parameter
to consider. As $\omega_{\rm DE}$ is constant in redshift
for a $\Lambda$CDM universe, the deviations in the value of this parameter recovered 
from different data sets could be used as an indication of their consistency within
the standard cosmological model. 

\section{Conclusions}
\label{section:conclusions}

We obtained constraints on the parameters of the standard $\Lambda$CDM model 
from the joint analysis of  anisotropic clustering measurements in configuration space
from BOSS and eBOSS. In particular, we used the information of the 
full shape of the clustering wedges of the final BOSS galaxy samples obtained 
by \citet{S17} and the legendre multipoles of the eBOSS DR16 QSO  
catalogue of \citet{Hou2021}. 
We updated the recipes to describe the non-linear matter power spectrum and the 
non-local bias parameters with respect to those used in the BOSS-only analyses of 
\citet{S17} and \citet{Tr_ster_2020}.
We directly compared our theoretical predictions for different cosmologies against the 
BOSS and eBOSS clustering measurements, without the commonly used RSD and 
BAO summary statistics.
We focus on cosmological parameters that can be classified 
either as shape or evolution parameters \citep{Sanchez2021}, such as the physical 
matter and dark energy densities, instead of other commonly used quantities 
such as $\Omega_{\rm m}$ and $\Omega_{\rm DE}$ that depend on the 
value of $h$. 
Our constraints from the combination of BOSS~+~eBOSS represent improvements ranging from 
20 to 25 per-cent with respect to those of \citet{Tr_ster_2020} and are in excellent agreement 
with Planck, with the suspiciousness and updated difference in means tension metrics 
indicating agreement at the level of 0.64$\sigma$ and  0.90$\sigma$, respectively. 

We combined the clustering data from BOSS and eBOSS with the $3\times 2$pt correlation 
function measurements from DES Y1 to obtain joint low-redshift cosmological constraints 
that are also consistent with the $\Lambda$CDM Planck results, 
albeit with larger deviations (1.54$\sigma$ and 2.00$\sigma$ differences as inferred 
from $S$ and $\mathcal{Q}_{\rm{UDM}}$, respectively). 
We do see interesting discrepancies in certain parameter combinations at the level of 
$2\sigma$ or more, such as the $\Omega_{\rm{m}}$ -- $\sigma_{8}$ and 
$\omega_{\rm{m}}$ -- $\sigma_{12}$ planes, and, 
more significantly, in the $\log(  10^{10}A_{\rm{s}})$ -- $\sigma_{12}$ projection. 
For a given value of $\sigma_{12}$, low-redshift probes (mostly driven by DES) prefer 
a higher amplitude of primordial density fluctuations than Planck, indicating differences 
in the total growth of structure predicted by these data sets. 

We further tested the impact of imposing a Gaussian prior on 
$\omega_{\rm b}$, $\omega_{\rm c}$, and $n_{\rm s}$ representing 
the constraints on these shape parameters recovered from Planck data.
Such prior leads to a significant improvement in the constraints on the evolution 
parameters, such as  $\omega_{\rm DE}$ and $A_{\rm s}$. 
In this case, we find that the full combination of low-redshift data sets prefers 
a value of the physical dark energy density $\omega_{\rm{DE}}$ that is 
1.7$\sigma$ higher than that preferred by Planck. 
This discrepancy, which is also related to the amount of structure growth preferred by 
these data sets, offers and interesting link with the $H_0$ tension, as it points to a higher value of 
$h$ being preferred by the low-redshift data.

The advent of new large, high-quality data sets such as the Dark Energy Spectroscopic 
Instrument \citep[DESI,][]{desi_survey}, the ESA space mission {\it Euclid} \citep{Laureijs2011}, 
and the Legacy Survey of Space and Time (LSST) at the Rubin Observatory \citep{Ivezic2019}, 
will allow us to combine multiple probes and significantly tighten our cosmological constraints. 
The discussion of the consistency between different data sets has so far been centred on 
the comparison of constraints on $S_8(z=0)$. 
As we move on to the analysis of Stage IV data sets, it would be beneficial to shift our 
focus from best constrained parameter combinations within a $\Lambda$CDM scenario to 
quantities that more closely represent the cosmological information content of those data, 
or that have a more direct physical interpretation. 

\section*{Acknowledgements}

We would like to thank Benjam\'in Camacho, Daniel Farrow, 
Martha Lippich, Tilman Tr\"oster, and Marco Raveri
for their help and useful suggestions. 
This research was supported by the Excellence Cluster ORIGINS, 
which is funded by the Deutsche 
Forschungsgemeinschaft (DFG, German Research Foundation) under 
Germany's Excellence Strategy - EXC-2094 - 390783311.

G.R. acknowledges support from the National Research Foundation of Korea (NRF) through Grants No. 2017R1E1A1A01077508 and No.2020R1A2C1005655 funded by the Korean Ministry of Education, Scienceand Technology (MoEST)

Funding for the Sloan Digital Sky Survey IV has been provided by the Alfred P. Sloan Foundation, the U.S. Department of Energy Office of Science, and the Participating Institutions. SDSS-IV acknowledges
support and resources from the Center for High-Performance Computing at
the University of Utah. The SDSS web site is www.sdss.org.

SDSS-IV is managed by the Astrophysical Research Consortium for the 
Participating Institutions of the SDSS Collaboration including the 
Brazilian Participation Group, the Carnegie Institution for Science, 
Carnegie Mellon University, the Chilean Participation Group, the French Participation Group, Harvard-Smithsonian Center for Astrophysics, 
Instituto de Astrof\'isica de Canarias, The Johns Hopkins University, Kavli Institute for the Physics and Mathematics of the Universe (IPMU) / 
University of Tokyo, the Korean Participation Group, Lawrence Berkeley National Laboratory, 
Leibniz Institut f\"ur Astrophysik Potsdam (AIP),  
Max-Planck-Institut f\"ur Astronomie (MPIA Heidelberg), 
Max-Planck-Institut f\"ur Astrophysik (MPA Garching), 
Max-Planck-Institut f\"ur Extraterrestrische Physik (MPE), 
National Astronomical Observatories of China, New Mexico State University, 
New York University, University of Notre Dame, 
Observat\'ario Nacional / MCTI, The Ohio State University, 
Pennsylvania State University, Shanghai Astronomical Observatory, 
United Kingdom Participation Group,
Universidad Nacional Aut\'onoma de M\'exico, University of Arizona, 
University of Colorado Boulder, University of Oxford, University of Portsmouth, 
University of Utah, University of Virginia, University of Washington, University of Wisconsin, 
Vanderbilt University, and Yale University.

Based on data products from observations made with ESO Telescopes at the La Silla Paranal Observatory under programme IDs 177.A-3016, 177.A-3017, 177.A-3018, 179.A-2004, and 298.A-5015.

\section*{Data availability}

The clustering measurements from BOSS and eBOSS used in this analysis are publicly available 
 via the SDSS Science Archive Server (https://sas.sdss.org/).



\bibliographystyle{mnras}
\bibliography{main} 



\appendix

\section{Joint analysis with KiDS-450 data}
\label{appendix:kids}

In this section we present the joint analysis of the anisotropic clustering 
measurements from BOSS and eBOSS together with cosmic shear measurements 
from KiDS-450, \citealp[][]{Hildebrandt_2016}. 

We use cosmic shear measurements from the Kilo-Degree Survey, \citep{Kuijken_2015, Hildebrandt_2016, Conti_2017}, hereafter referred to as KiDS. The KiDS data are processed by THELI \citep{Erben_2013} and Astro-WISE \citep{Begeman_2013, deJong_2017}. Shears are measured using lensfit \citep{Miller_2013}, and photometric redshifts are obtained from PSF-matched photometry and calibrated using external overlapping spectroscopic surveys (see \citet{Hildebrandt_2016}).

The KiDS-450 weak lensing data set consists of tomographic shear measurements from four  
redshift bins spanning the total range of $0.1<z\leq0.9$ and the corresponding source redshift 
distributions estimated from the weighted direct calibration (`DIR') for each bin \citep{Lima_DIR}. 
We use the recommended scale cuts and  use 
the angular bins with $\theta < 72\,{\rm arcmin}$ for $\xi_+(\theta)$ and $\theta > 6\,{\rm arcmin}$ 
for $\xi_-(\theta)$. 

\begin{figure}
\centering
	\includegraphics[width=0.8\columnwidth]{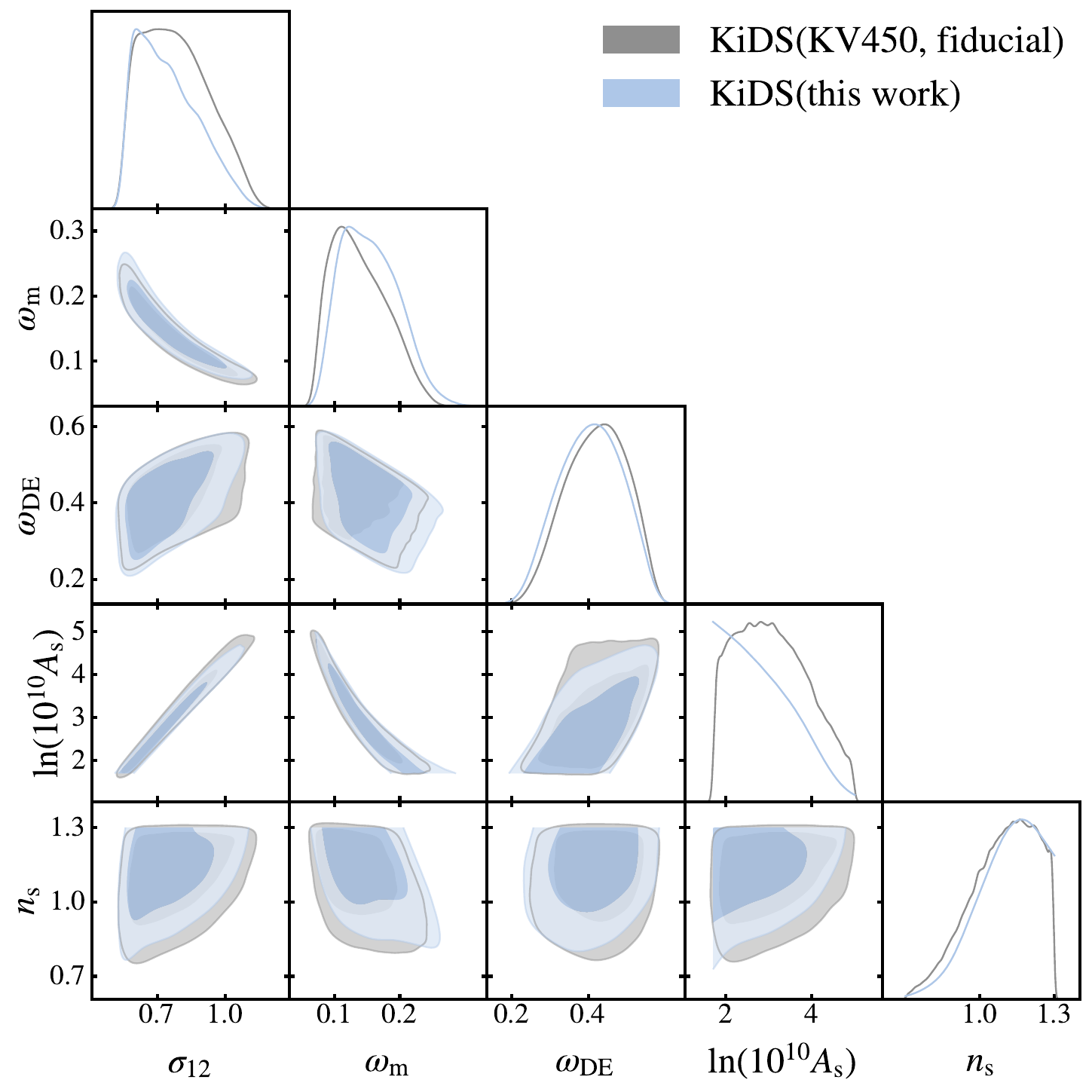}
    \caption{Comparison of the marginalised posterior distributions between the fiducial KV450 
    analysis and this work. For this comparison, we adapted the cosmological parameter priors 
    to match those of the fiducial analysis.}
    \label{fig:kids_validate}
\end{figure}

It is important to note that here we are using the same DES shear model as in the main analysis. 
This means that the treatment of the nuisance parameters (namely, the baryonic effects, the photometric redshift 
uncertainty and the additive and multiplicative bias parameters) differs from the original analysis of KiDS-450. We do not include baryonic effects and our priors for photometric redshift uncertainty match those of \citet{kidsviking}. We impose flat priors for multiplicative bias $U(-0.1,0.1)$ with the additive bias parameter taken to be zero, mimicking the DES set up. Finally, we also follow the DES intrinsic alignment model, imposing a flat prior on redshift evolution of intrinsic alignment parameter.
We compare our final posteriors with the ones obtained from the publicly available 
KV450 chains \citep{kidsviking} and find a good agreement between the two, as shown in 
Fig.~\ref{fig:kids_validate}. Moreover, there is a weak tendency is for our analysis to prefer lower 
RMS variance values, which leads to a more conservative assessment of any potential tensions with Planck. 

\begin{figure}
\centering
	\includegraphics[width=0.8\columnwidth]{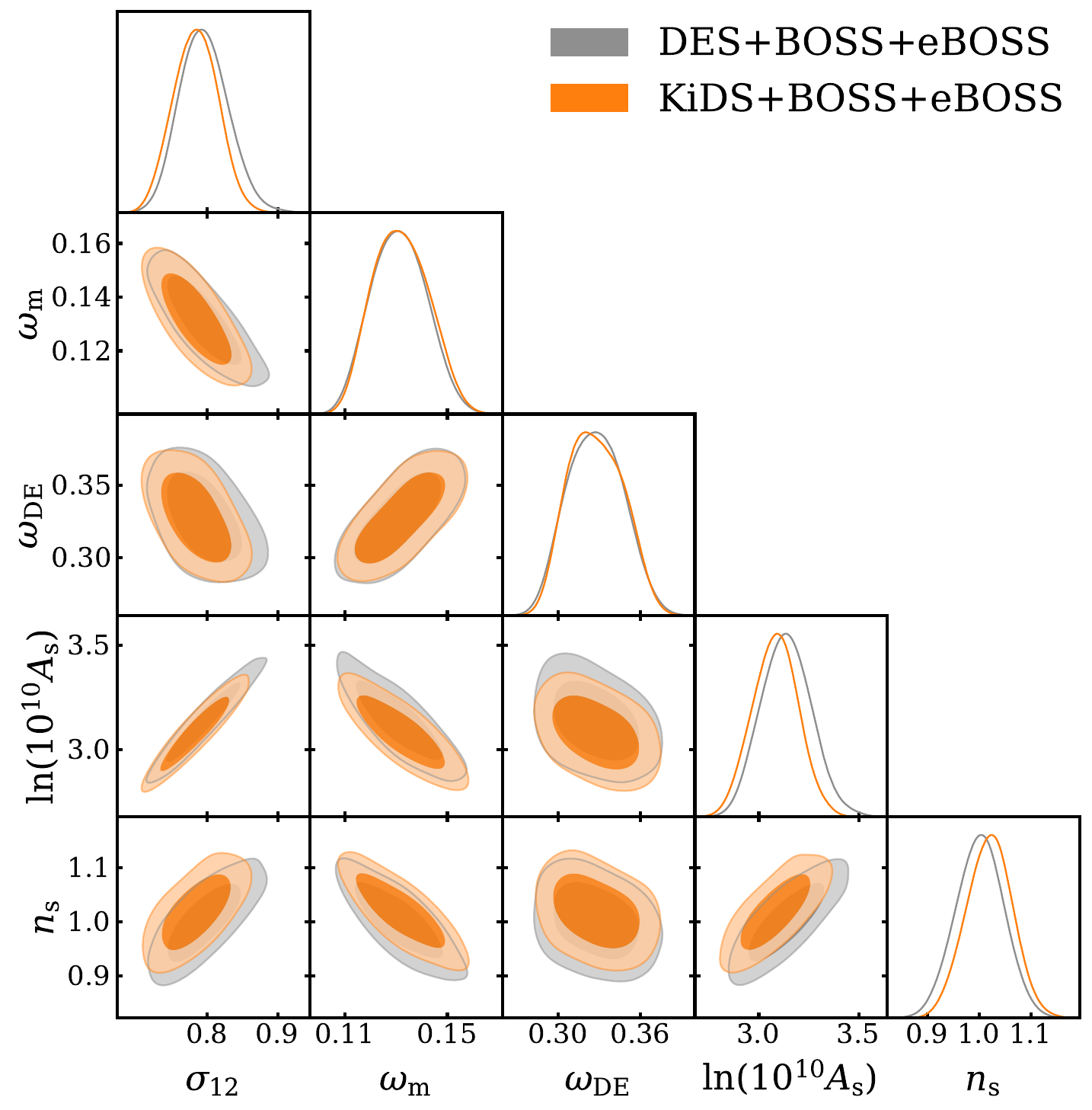}
    \caption{Comparison of low-redshift constraints obtained by combining BOSS~+~eBOSS with 
    either DES Y1 $3\times 2$pt data (grey contours) or the KiDS-450 shear 
    measurements (orange contours).}
    \label{fig:kidsvsdes}
\end{figure}

Fig.~\ref{fig:kidsvsdes} shows the combined constraints from the combination of 
KiDS-450~+~BOSS~+eBOSS (orange contours), which are in near-perfect agreement 
with the equivalent combination using DES (grey contours). The suspiciousness statistic shows 
an agreement between Planck and this set of  low-redshift measurements of 
$1.5 \pm 0.5 \sigma$, which is also consistent with our results from BOSS~+~eBOSS~+~DES. 

\bsp	
\label{lastpage}
\end{document}